# How journal rankings can suppress interdisciplinary research: A comparison between Innovation Studies and Business & Management[1]

Ismael Rafols[1,2], Loet Leydesdorff[3], Alice O'Hare[1], Paul Nightingale[1] and Andy Stirling[1]

[1] i.rafols@sussex.ac.uk (Corresponding author)
SPRU —Science and Technology Policy Research, University of Sussex, Brighton, BN1 9QE, England

[2] Technology Policy and Assessment Center, Georgia Institute of Technology, Atlanta, GA 30332, USA

[3] loet@leydesdorff.net
University of Amsterdam, Amsterdam School of Communication Research (ASCoR)
1012 CX Amsterdam, The Netherlands

**Abstract**

This study provides quantitative evidence on how the use of journal rankings can disadvantage interdisciplinary research in research evaluations. Using publication and citation data, it compares the degree of interdisciplinarity and the research performance of a number of Innovation Studies units with that of leading Business & Management schools in the UK. On the basis of various mappings and metrics, this study shows that: (i) Innovation Studies units are consistently more interdisciplinary in their research than Business & Management schools; (ii) the top journals in the Association of Business Schools' rankings span a less diverse set of disciplines than lower-ranked journals; (iii) this results in a more favourable assessment of the performance of Business & Management schools, which are more disciplinary-focused. This citation-based analysis challenges the journal ranking-based assessment. In short, the investigation illustrates how ostensibly 'excellence-based' journal rankings exhibit a systematic bias in favour of mono-disciplinary research. The paper concludes with a discussion of implications of these phenomena, in particular how the bias is likely to affect negatively the evaluation and associated financial resourcing of interdisciplinary research organisations, and may result in researchers becoming more compliant with disciplinary authority over time.

---

[1] This article received the William Page Best Paper Award in the 2011 *Atlanta Science and Innovation Policy Conference*.

**Highlights**

- We compare Innovation Studies units (IS) with Business & Management schools (BMS).
- IS are found to be more interdisciplinary than BMS according to various metrics.
- BMS have higher performance according to indicators based on journal rankings.
- This higher performance of BMS is not supported by citation-based indicators.
- The analysis suggests that journal rankings are biased against interdisciplinarity.

**1. Introduction**

At a time when science is under pressure to become more relevant to society (Nightingale and Scott, 2007; Hessels, 2010), interdisciplinary research (IDR) is often praised for contributing to scientific breakthroughs (Hollingsworth and Hollingsworth, 2000), for addressing societal problems (Lowe and Phillipson, 2006) and for fostering innovation (Gibbons et al., 1994). Reasons given for supporting IDR include suggestions that it is better at problem-solving (Page, 2007, p. 16), that it generates new research avenues by challenging established beliefs (Barry et al., 2008), and that it is a source of creativity (Heinze et al., 2009; Hemlin et al., 2004). These are all claimed to help rejuvenate science and contribute towards its ongoing 'health' (Jacobs and Frickel, 2009, p. 48).

However, IDR is also widely perceived as being at something of a disadvantage when it comes to research evaluation (Rinia et al., 2001a, p. 357; Nightingale and Scott, 2007, pp. 546-547). Various qualitative studies have provided evidence that peer review tends to be biased against IDR (Laudel and Origgi, 2006; Langfeldt, 2006, p. 31). However, only a few quantitative investigations have been undertaken of this claim, and they have been mostly inconclusive (Porter and Rossini, 1985, p. 37; Rinia et al., 2001a).

Here we explore potential biases in the evaluation of IDR in the particular case of Innovation Studies (IS) units in the UK. Innovation Studies is a diverse and rather ambiguously bounded area of social science that studies the causes, processes and consequences of innovation (Fagerberg et al., this issue). Given its problem-oriented and interdisciplinary nature, Innovation Studies research is conducted in diverse types of research units that experience a variety of institutional challenges (Clausen et al., this issue), in particular a lack of fit with discipline-based assessment panels.

The UK is a particularly suitable setting for this enquiry, as it has a sizeable and well established IS community, a comparatively homogenous higher education system, and a long history of research assessment (Collini, 2008). The UK has also witnessed repeated concerns about possible biases against IDR – not least following the Boden Report (ABRC, 1990). Under the funding conditions prevailing in the UK, many IS units have in recent years been (at least partly) incorporated into, or linked with, Business and Management Schools (BMS) (e.g. in Oxford, Imperial, Manchester, Cardiff and recently Sussex). BMS face acute pressures to achieve high performance in institutional rankings, both for reputational purposes and because of the financial incentives associated with the research assessment procedures of the UK's national funding council, HEFCE.[2] This assessment exercise (which was formerly known as the research assessment exercise or RAE) is currently referred to as the 'Research Excellence Framework' (REF) (Martin and Whitley, 2010, p. 61). BMS in the UK are also subject to a narrowly-conceived formal ranking scheme for journals, provided by the British Association of Business Schools (ABS) (ABS, 2010).

---

[2] Higher Education Funding Council for England.

The use of journal rankings[3] (such as those provided by ABS) in research evaluations has become increasingly popular. It is seen as a means to 'objectify' research assessment and thus avoid or compensate for any biases in peer review (Taylor, 2011). Yet journal-based evaluation has been severely criticised as being inappropriate for this role (Seglen, 1997; Oswald, 2007). Despite this, the proliferation of journal ranking schemes indicates increasingly wide usage across disciplines (both explicitly and implicitly) for a variety of quality assessment purposes, such as resourcing, recruitment and promotion. A range of studies have demonstrated that the journal ranks of a department's publications are by far the strongest predictor of the results obtained in the 2008 UK's RAE, although journals rankings were not formally used in the evaluation (Kelly et al., 2009; Taylor, 2011, pp. 212-214). As a result, university managers are making increasingly explicit use of such journal rankings to prepare future assessments.

In this study, three centres for IS in the UK are compared with the three leading British BMS. The choice of BMS as comparators is influenced by the fact that many IS centres are now closely associated with BMS and hence will be assessed by the Business & Management panel in the forthcoming REF. We investigate quantitatively the relationship between the degree of interdisciplinarity and perceived performance, as shown by the ABS journal rankings. We then compare the results with arguably more reliable article-based citation indicators. In summary, the results suggest that ABS journal rankings favour research within the dominant disciplines of BMS (mainly business, management, economics and finance) and disadvantage interdisciplinary IS units. Given the close correlation between RAE grades and assessments based on journal ranks in previous RAEs (Taylor, 2011), this effect is large enough to have a substantial negative impact on the funding of IS units.

This study makes two contributions. First, it is (to our knowledge) the first to demonstrate a bias against IDR on a firm quantitative basis (Porter and Rossini, 1985, p. 37; Rinia et al., 2001a). Second, it shows that bias against IDR may arise not only in peer review – as well documented by qualitative studies (Laudel and Origgi, 2006) – but also in purportedly *objective* assessment, such as quantitative journal rankings.

The policy implications of these results will be discussed in the light of studies on the consequences of biases in assessments. For example, research suggests that British economics departments have narrowed their recruitment to favour 'main-stream' economists (Harley and Lee, 1997; Lee and Harley, 1998; Lee, 2007), thus reducing the cognitive diversity of the research system's ecology. This may lead to intellectual impoverishment in the medium or long term (Molas-Gallart and Salter, 2002; Stirling, 1998, pp. 6-36; Stirling, 2007; Martin and Whitley, 2010, pp. 64-67).

In addition to its primary focus on the bias against IDR in research assessment, this article also aims to make a more general contribution to advancing the state-of-the-art with regard to the use of bibliometric indicators for policy purposes. First, it provides an introduction to a variety of concepts, mathematical operationalisations and visualisations for the study of interdisciplinarity using bibliometric data. Second, it highlights that conventional measures of performance for IDR publications remain problematic, and suggests 'citing-side normalisation' as an improved alternative. Third, it illustrates the use of multiple indicators for the study of multidimensional concepts such as interdisciplinarity or research performance. In this, we follow Martin and Irvine's (1983) seminal argument that, since no simple measures exist that can fully capture the research contributions made by scientists, one should use various partial indicators. Though incomplete (as well as being imperfect and subject to contingency and distortion), this more 'plural and conditional' (Stirling, 2010) form of bibliometric analysis may be considered to be more reliable when diverse indicators converge to yield broadly the same insights. Since plurality is more easily captured by multidimensional representations, we illustrate this point with a full set of maps (available at http://interdisciplinaryscience.net/maps/ and in the supplementary materials).

---

[3] Most so-called 'journal rankings' provide a 'rating' of a large list of journals. Since in some cases, the ratings are integer numbers, they do not necessarily constitute an absolute 'ranking' (i.e. ordering) of the journals. However, since the main objective of the exercise is to compare and select journals by classifying them into 'ranks', we keep in this article the term 'journal ranking', in agreement with most science policy documents and academic literature.

For the sake of focus, a number of otherwise relevant issues related to the subject will not be dwelt on in this article. In particular, the present study does not offer any kind of assessment of the individual organisations examined – this would entail a broader evaluation than the exclusive focus on publication output and impact used here. Second, it does not discuss the relative benefits of IDR. We simply note that IDR is highly valued by many researchers and policy-makers – which is sufficient to render important the question of whether IDR is fairly assessed. Third, we do not look into the broader societal impact of research. The concern here is whether there is a bias against IDR only when considering conservative, internal measures of scientific merit. Finally, we do not elaborate the details of conceptualisations and operationalisations of interdisciplinarity and performance. Instead, we build on fairly conventional indicators of performance and on published research on IDR. Given the length of the paper, some readers may prefer to skip section 2 (literature review), section 3 (data and methods), and section 5 (discussion), and concentrate their attention on section 4 (results) and section 6 (conclusions), before returning to the rest of the paper.

## 2. The evaluation of interdisciplinarity research

Various notions of interdisciplinarity have become prominent in science policy and management (Metzger and Zare, 1999). IDR is seen as a way of sparking creativity, supporting innovation and addressing pressing social needs (Jacobs and Frickel, 2009, p. 48). This is well-illustrated by a variety of high profile initiatives, such as the UK's Rural Economy and Land Use Programme (RELU[4], Lowe and Phillipson, 2006), the US Integrative Graduate Education and Research Traineeship (IGERT,[5] Rhoten et al., 2009), the explicit call to cross disciplinary boundaries in the prestigious grants of the new European Research Council (ERC, 2010, p.12), or the establishment of new cross-disciplinary institutes such as the Janelia Farm of the Howard Hughes Medical Institute or the Bio-X centre at Stanford University (Cech and Rubin, 2004). These developments have been accompanied by significant increases in articles claiming to be interdisciplinary (Braun and Schubert, 2003) and by a shift towards more interdisciplinary citing patterns (Porter and Rafols, 2009).

However, in parallel with this wave of declared support, IDR is, in practice, often accused of being too risk-averse, of lacking in terms of disciplinary notions of quality, or of not meeting policy expectations (Bruce et al., 2004, pp. 468-469). Claims over the benefits of interdisciplinarity are questioned (Jacobs and Frickel, 2009, p. 60), since they are often based on limited evidence relying heavily on 'cherry-picked' case-studies of success that have been selected and analysed ex-post (e.g. Heinze et al., 2009 or Hollingsworth and Hollingsworth, 2000). The effects of IDR on research outcomes are difficult to prove systematically because interdisciplinarity is just one of many mediating factors that contribute to the success or relevance of research. As a result, subtle contextual differences can lead to disparate results. For example, whereas some studies have correlated IDR practices with the intensity of university-industry interactions (Van Rijnsoever and Hessels, 2011; Carayol and Thi, 2005), other studies do not find that IDR influences the success of firms founded by academic teams (Muller, 2009).

---

[4] http://www.relu.ac.uk/
[5] http://www.igert.org/

Yet, irrespective of the perspective adopted, there is agreement that IDR faces important barriers that may significantly hinder its potential contributions (Rhoten and Parker, 2006; Llerena and Mayer-Krahmer, 2004). In the first place there are difficulties in managing the coordination and integration of distributed knowledge. This has been addressed by research examining various kinds of team work and collaboration (Cumming and Kiesler, 2005, 2007; Katz and Martin, 1997; Rafols, 2007).[6]

Second, there are more systemic barriers stemming from the institutionalisation of science along disciplinary lines (Campbell, 1969; Lowe and Phillipson, 2009). Perceived barriers include the relatively poor career prospects often experienced by interdisciplinary researchers, lower esteem from colleagues, discrimination by reviewers in proposals, and disproportionate difficulty in publishing in prestigious journals (Bruce et al., 2004, p. 464). The US National Academies (2004) report on *Facilitating Interdisciplinary Research* provides a thorough review of these barriers, and suggests various initiatives to lower them. Since these barriers tend to be embedded and thus implicitly 'naturalised' in institutional practices, they are generally less visible and more controversial than teamwork problems. While such hurdles for IDR are often acknowledged in policy initiatives, the mechanisms by which they operate are neither well documented nor clearly understood (EURAB, 2004; Metzger and Zare, 1999; National Academies, 2004; Rhoten and Parker, 2006).

One widely perceived 'key barrier' is the apparent bias against IDR in most research evaluation (Rinia et al., 2001a, p. 357; Lee, 2006; Nightingale and Scott, 2007, pp. 546-547). For example, Boddington and Coe (1999, p.14) reported from a large survey (of 5,505 respondents) that 51% of researchers, 68% of department heads and 48% of RAE panel members viewed the 1996 UK RAE as slightly or strongly inhibiting IDR (as compared to 24%, 15% and 19%, respectively, who saw RAE as promoting IDR). Investigations on peer-review-based research evaluation support these perceptions (see e.g. the special issue of the journal *Research Evaluation* edited by Laudel and Origgi, 2006). In summary:

> '...a re-emerging awareness of interdisciplinarity as a vital form of knowledge production is accompanied by an increasing unease about what is often viewed as the 'dubious quality' of interdisciplinary work. Central to the controversy is the lingering challenge of assessing interdisciplinary work.' (Boix Mansilla, 2006, p.17)

---

[6] An important research agenda focusing on these problems in recent years is associated with the so-called 'Science of Team Science' (SciTS) community, which has developed 'an amalgam of conceptual and methodological strategies aimed at understanding and enhancing the outcomes of large-scale collaborative research and training' (Stokols et al., 2008. p. S77; Börner et al., 2010).

That evaluation of IDR is problematic is not a surprise. Any evaluation needs to take place using established standards. These standards can be defined within a narrow discipline, but what standards should be used for research in between or beyond existing disciplinary practices? If IDR must meet the (sometimes radically) contrasting quality criteria of more than one discipline, then it self-evidently faces an additional hurdle, compared to mono-disciplinary research which is evaluated against a single set of criteria. Beyond this, peer-review has been shown to exhibit inherently conservative and risk-minimising tendencies, which 'may disfavour unconventional and interdisciplinary research' (Langfeldt, 2006, p. 31) and hence favour well established fields over nascent ones (Porter and Rossini, 1985, p. 37). Of course, programmes targeting 'high risk, high reward research', where IDR is explicitly encouraged, can be an exception to this (Balakrishnan et al., 2011). But what appears to happen generally, even in the case of multidisciplinary review panels, is that IDR ends up being assessed from the perspective of what appears to be the most relevant discipline (Mallard et al., 2009, p. 22) or under the evaluator's own favoured disciplinary criteria, a phenomenon dubbed 'cognitive cronyism or particularism' (Travis and Collins, 1991).[7] As Laudel and Origgi (2006, p. 2) note:

> 'in spite of the political narratives on the need for interdisciplinarity, the criterion of quality can be turned into an instrument for suppressing interdisciplinary research because the established [disciplinary] quality standards are likely to prevail.'

Perhaps surprisingly, the strong impression from qualitative studies that IDR is at a disadvantage in peer review has apparently not been robustly substantiated by quantitative studies. Examining a total of 257 reviews of 38 projects from five somewhat interdisciplinary programmes (e.g. neurobiology) of the US National Science Foundation, Porter and Rossini (1985) found a weak but significant correlation between low grades and degree of interdisciplinarity ($r=0.29$, $p<0.05$). In contrast, Rinia et al. (2001a), who analysed the evaluation by an international panel of 185 physics programmes in Dutch universities, did not find a bias against IDR. However, they did note that IDR tends to be published in journals with a lower citation impact (Rinia et al., 2001a, p. 360; 2001b, p. 247).

In the previously mentioned survey commissioned by the UK Higher Education Funding Council on the RAE, Boddington and Coe (1999, p.iii) concluded that 'there is no evidence that the RAE systematically discriminated against interdisciplinary research in 1996'. Interestingly, though, a closer look at their data shows that the highest RAE scores were obtained by researchers that reported themselves as being at the lower end of the interdisciplinary spectrum (dedicating between 30-40% of their time to IDR activities), whereas researchers reporting high IDR-involvement obtained lower scores. The effect is particularly strong for teaching-based researchers.[8] Other bibliometric studies have found that articles with an intermediate degree of interdisciplinarity are more likely to be cited than either the mono-disciplinary or the extremely interdisciplinary ones (e.g. Adams et al., 2007; Larivière and Gingras, 2010; Yegros-Yegros et al., 2010), and that in the natural sciences the average number of citations per paper received by multidisciplinary journals[9] is lower than in

---

[7] 'Cognitive particularism' should not be confused with the better documented phenomenon of institutional capture by 'old boys' networks.

[8] Although the study does not report any standard error or statistical significance of the results, one might assume that the results are likely to be statistically significant since they are based on a sample of 5,505 respondents.

[9] Multidisciplinary journals are defined by Levitt and Thelwall as journals that are classified into two or more subject categories. Journals satisfying this definition play a bridging role between disciplines. Notice that this is

mono-disciplinary ones (Levitt & Thelwall, 2008). However, since these studies did not make comparisons between bibliometric data and research evaluation rankings, any potential biases against IDR could not be assessed.

In summary, in contrast to the numerous qualitative studies pointing to clear bias against IDR in evaluation (Travis and Collins, 1991; Langfeldt, 2006), there are only a few quantitative studies on the subject and these have produced ambiguous and somewhat contradictory results. This overall inconclusiveness in quantitative evidence has been interpreted by some as evidence for the absence of bias (Huutoniemi, 2010, p. 318). In this study we aim to help fill this gap by investigating a potential bias against IDR that results from the use of journal rankings in research evaluation.

### 3. Methods and underlying conceptualisations

*3.1 Methodological framework: converging partial indicators*

Assessments of scientific performance and interdisciplinarity remain controversial and exhibit no consensus on appropriate frameworks and methodologies, even when based on narrow quantitative measures such as publication outputs (Bordons et al., 2004; Huutoniemi et al., 2010). This should come as no surprise, given that both performance and interdisciplinarity are essentially multidimensional concepts, which can only be partially captured by any single indicator (Martin and Irvine, 1983; Narin and Hamilton, 1996; Sanz-Menéndez et al., 2001).

Unfortunately, as scientometrics became more widely used and institutionalised in policy and management, flaws (and associated caveats – see e.g. Leydesdorff, 2008) in the use of bibliometric tools have become increasingly overlooked. Martin (1996) reminded the research policy community of the lack of robustness of one-dimensional measurements of multi-dimensional concepts such as interdisciplinarity or scientific performance. Particularly under conditions of uncertainty and ambiguity, there is a need for more 'plural and conditional' metrics (Stirling, 2010) if research assessments are to become more accurate and reliable.

Here we follow the main tenets of the 'converging partial indicators' method (Martin and Irvine, 1983) and enlarge its scope by using recently developed mapping techniques, which help end-users explore their own partial perspectives by providing them with a range of diverse indicators. Not only is this approach more robust, it is also more likely to be recognised as legitimate when the diverse perspectives converge on a similar conclusion. This is arguably the case for the findings on interdisciplinarity presented here. Alternatively, when different approaches lead to contradictory insights, it becomes clear that the conclusions are more questionable, possibly reflecting the choice of indicator as much as the phenomenon under investigation. As we shall see, this is arguably the case for the findings presented here specifically on performance. In order for the reader to be able to engage in this exploration, the full set of 54 maps (9 for each organisation) used for the analysis is available at http://interdisciplinaryscience.net/maps/ and in the supplementary materials.

*3.2 The assessment of interdisciplinarity*

---

different from the popular understanding of 'multidisciplinary' journals, such as *Nature* and *Science*, which publish articles from several disciplines (but mostly mono-disciplinary ones) for a wide audience.

The inherently ambiguous, plural and controversial features of prevailing understandings of interdisciplinarity have inevitably led to a lack of consensus on indicators (see Wagner et al., 2011, for a review). Even within bibliometrics, the operationalisation of IDR remains contentious and defies one-dimensional descriptions (Bordons et al., 2004; Huutoniemi et al., 2010; Leydesdorff and Rafols, 2011a; Sanz-Menéndez et al., 2001). We propose to investigate interdisciplinarity from two perspectives. The first is by means of the widely-used conceptualisation of interdisciplinarity as *knowledge integration* (National Academies, 2004; Porter et al., 2006), which is perceived as crucial for innovation or solving social problems. The second is by conceptualising interdisciplinarity as a form of research that lies outside or in between established practices, i.e. in terms of *intermediation* (Leydesdorff, 2007a).

Understanding interdisciplinarity as integration suggests looking at the distribution of components (disciplines or sub-disciplines) that have been linked or integrated under a body of research (as shown by a given output, such as a reference list). We do so here by using the concepts of diversity and coherence, as illustrated in Figure 1 (Rafols and Meyer, 2010).[10] We propose to explore knowledge integration in two steps. The first involves employing the concept of *diversity* as 'an attribute of any system whose elements may be apportioned into categories' (Stirling, 2007, p. 708). This allows exploration of the distribution of disciplines to which parts of a given body of research can be assigned.

A review of the literature reveals that many bibliometric and econometric studies of interdisciplinarity were based on (rather incomplete, as we will later see) indicators of diversity such as Shannon entropy (Carayol and Thi, 2005; Hamilton et al., 2005; Adams et al., 2007) and Simpson diversity (equivalent to the Herfindahl index in economics, and often used in patent analysis – see e.g. Youtie et al., 2008). However, knowledge integration is not only about how diverse the knowledge is, but also about making connections between the various bodies of knowledge drawn upon. Hence, the second step for the operationalisation of integration is assessing the extent to which distant disciplines in the case under study are linked – something that we explore here with the concept of *coherence*.

Although not a central focus of this article, we briefly note that the two-dimensional matrix of diversity vs. coherence shown in Figure 1 offers a systematic ordering of differences between mono-disciplinary, multidisciplinary and interdisciplinary research (National Academies, 2004, pp. 27-29).[11]

Understanding interdisciplinarity in terms of intermediation was first proposed by Leydesdorff (2007a), building on the concept of 'betweenness centrality' (Freeman, 1977). As illustrated in Figure 2, intermediation does not entail combining diverse bodies of knowledge, but contributing to a body of knowledge that is not in any of the dominant disciplinary territories. As in the case shown in the right hand side of Figure 2, even when diversity is low, a case can be considered interdisciplinary if a large proportion of its components are in intermediate positions.

---

[10] See a more general conceptual framework developed in Liu et al. (2012).
[11] The fourth quadrant in the Coherence vs. Diversity scheme would be a possibly unusual case of research that is focused and that it nevertheless exhibits relatively strong connections with disparate fields. We tentatively call such cases 'reflexively disciplinary'.

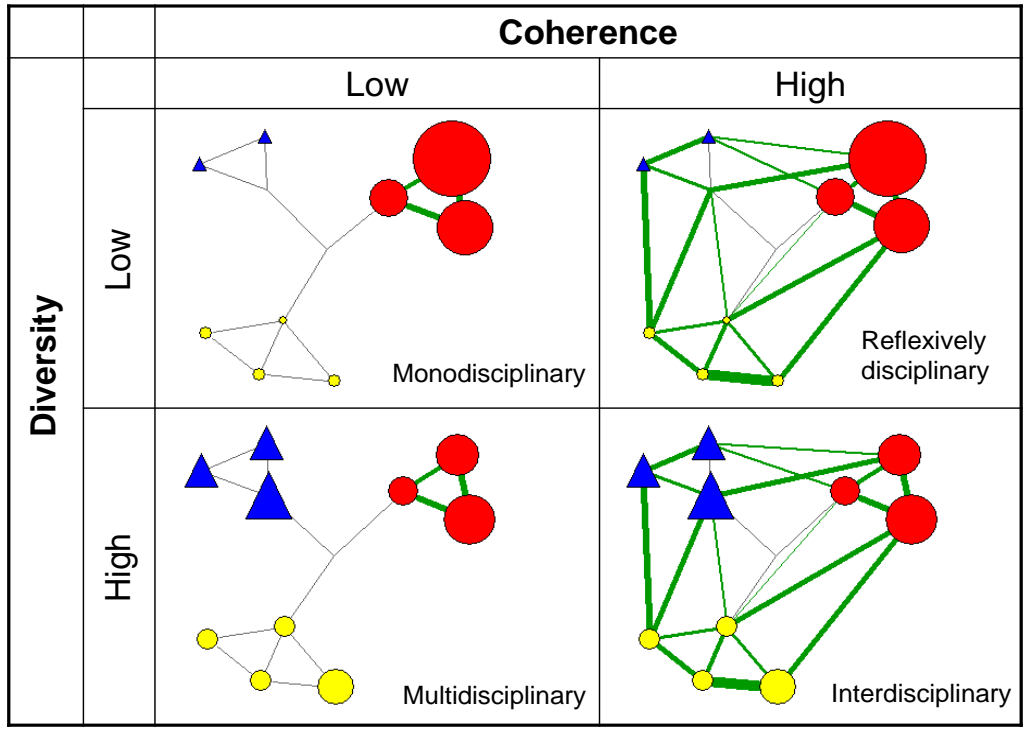

**Figure 1. Conceptualisation of interdisciplinarity in terms on knowledge integration**
Each node in the networks represents a sub-discipline. Grey lines show strong similarity between sub-disciplines. Same colours (shapes) illustrate clusters of sub-disciplines forming a discipline. Green lines represent direct interaction between sub-disciplines. The size of nodes portrays relative activity of an organisation in a given sub-discipline. Knowledge integration is achieved when an organisation is active in diverse sub-disciplines and interlinks them.

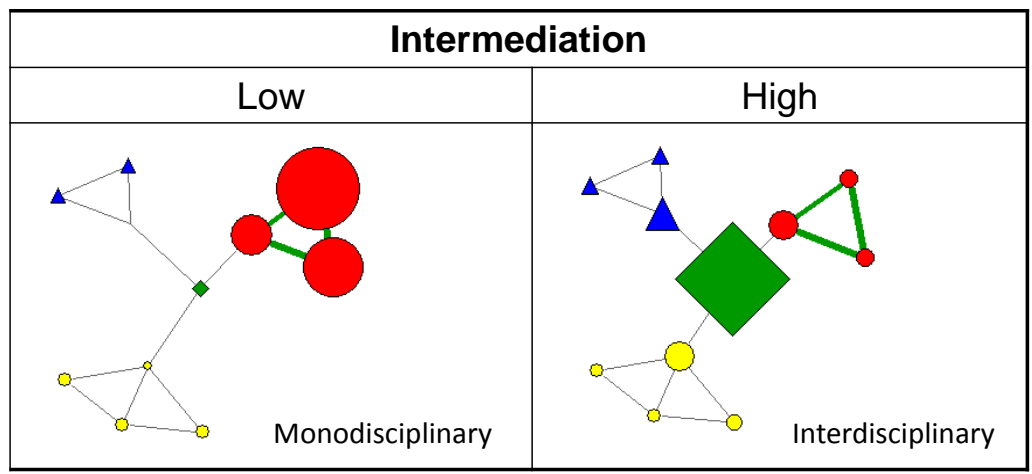

**Figure 2. Conceptualisation of interdisciplinarity as intermediation**
Intermediation is achieved when on organisation is active in disciplines or sub-disciplines (here the green rhomb) that occupy an interstitial position, i.e. between other disciplines (here the red or yellow circles and the blue triangles). See Figure 1 for further explanation of symbols.

A comparison between Figures 1 and 2 illustrates that knowledge integration and intermediation are two distinct processes. Although these properties may overlap, they do not need to occur at the same time. Indeed in a study on multiple measures of interdisciplinarity of journals, Leydesdorff and Rafols (2011a) found that they constituted two separate dimensions when using factor analysis.

Knowledge integration, on the one hand, occurs in research that builds on many different types of expertise. This is typically the case in emergent areas that combine disparate techniques from various fields, for example in medical applications of the 'lab on a chip', which draw both on micro-fabrication and biomedical expertise (Rafols, 2007). Intermediation, on the other hand, occurs when research does not readily fit with dominant disciplinary structures. This is often the case for instrumental bodies of knowledge, such as microscopy or statistical techniques, each with their own independent expertise, yet at the same time providing a service contribution to different disciplines (Price, 1984; Shinn and Joerges, 2002). Intermediation may also show up in what Barry et al. (2008, p. 29) called 'agonistic research', which emerges in opposition to the intellectual, ethical or political limits of established disciplines. Such research tends to push towards fragmentation, insularity and plurality rather than integration (Fuchsman, 2007). As a result, it is seldom captured in conventional classification categories. We therefore investigate intermediation at a lower level of aggregation than diversity and coherence.

We now describe how the concepts of diversity, coherence and intermediation can be operationalised. One advantage of using general concepts rather than ad hoc indicators is that it allows rigorous and plural comparison of – and choice between – different mathematical forms that are equally consistent with the processes we are seeking to capture. Hence the analysis follows the tenets of the 'converging partial indicators' approach (Martin and Irvine, 1983). The emphasis is not simply on the incidental value of multiple indicators or their collective ranges of variability (Funtowicz and Ravetz, 1990). The aim is also to focus deliberate, self-conscious and critical attention on the specific *conditions* under which different metrics (and their associated findings) are best justified (Stirling, 2008).

It has been widely documented across a diverse range of areas of appraisal that there are often strong institutional pressures artificially to reduce appreciations of uncertainty and complexity in evaluation, in order to justify particular favoured interpretations (Collingridge, 1982). It is in light of this problem that we deliberately use a 'plural and conditional' framework (rather than a multiplicity of indicators), in order to increase the accuracy and robustness of policy appraisal (Stirling, 2010). By explicitly discriminating between multiple, contrasting quantitative characterisations of disciplinary diversity, coherence and intermediation (each with its associated rationale and applicability), we can better document the specific phenomena under scrutiny, and also contribute methodologically towards the general ends of addressing bias and ensuring legitimacy when using scientometric indicators.

*Diversity*

A given body of research (as represented, for example, in the publications of a university department), can be considered to be more interdisciplinary if that department publishes in diverse disciplinary categories and the publications are coherent in the sense of linking the various categories. Diversity is a multidimensional property that has three main attributes (Stirling, 1998; 2007): *variety,* the number of categories of elements, in this case, the sub-disciplines into which publications can be partitioned; *balance,* the distribution across these categories, in this case, of output publications, or references in, or citations to, these (see details in methods, below); and *disparity*, the degree of distinctiveness between categories, in this case, the cognitive distance between sub-disciplines as measured by using bibliometric techniques (Leydesdorff and Rafols, 2009).

Figure 3 makes use of a specific type of science map, an 'overlay map', which is particularly appropriate for capturing diversity (Rafols et al., 2010, p. 1883). The overlay technique displays two key parameters. First, a baseline map shows the relations between elements in a large reference set of scientific activities (in this case, citations between sub-disciplines for the full Web of Science in 2009). Second, the relevant nodes in this baseline map are 'overlaid' with circles whose relative size represents the frequency of activity in each area of the particular subset under study (in this case a University unit).

An overlay representation of publication frequencies in the global map of science captures the three attributes of diversity. It shows whether the publications (or references or citations) of a department are dispersed over many or a few sub-disciplines (variety), whether the frequencies are evenly distributed (balance) and whether they are associated with proximate or distant areas of science (disparity). Since this is an inherently multidimensional description, in order to obtain scalar indicators one either has to consider each of the attributes separately or devise an indicator comprising the three aspects that makes a specific choice regarding the particular emphasis given to each attribute (variety, balance or disparity) (Stirling, 2007). Most previous studies on interdisciplinarity used indicators that rely on variety or balance (e.g. Larivière and Gingras, 2010), or combinations of both such as Shannon entropy (e.g. Carayol and Thi, 2005; Adams et al., 2007), but crucially missed taking into account the disparities among categories. In doing so, they implicitly consider as equally interdisciplinary a combination of cell biology and biochemistry (two related fields) and a combination of geology and psychology (two disparate fields). Only recently have new indicators incorporating disparity been devised, using the metrics of similarity behind the science maps (Porter et al., 2007; Rafols and Meyer, 2010). This operationalisation of diversity also allows us to visualize processes of knowledge diffusion (rather than integration) by looking at the disciplinary distribution of citations to a topic or an organisation's papers (Liu et al., 2012).[12]

Following Yegros-Yegros et al. (2010), we employ indicators that explore each of the dimensions separately and in combination. As a metric of distance we use $d_{ij} = 1 - s_{ij}$ with $s_{ij}$ being the cosine similarity between Subject Categories $i$ and $j$ (the metrics underlying the global science maps), and with $p_i$ being the proportion of elements (e.g. references) in category $i$. We explore the following indicators of diversity:

*Variety* (number of categories) $\quad n$

*Balance* (Shannon evenness) $\quad -\frac{1}{\ln(n)} \sum_i p_i \ln p_i$

*Disparity* (average dissimilarity between categories) $\quad \frac{1}{n(n-1)} \sum_{i,j} d_{ij}$

*Shannon entropy* $\quad -\sum_i p_i \ln p_i$

*Rao-Stirling diversity* $\quad \sum_{i,j} p_i p_j d_{ij}$

<u>Coherence</u>

---
[12] See also Kiss et al. (2010) and Leydesdorff and Rafols (2011b).

The term coherence refers to the extent to which the categories are connected to one another *within* the subset under study. Whereas measures of diversity are well established, measures of coherence (and intermediation) are still at an exploratory stage.[13] Here, to capture coherence we compare the *observed* average distance of cross-citations as they actually occur in the publications in question ($\sum_{i,j} p_{ij} d_{ij}$), with the *expected* average distance ($\sum_{i,j} p_i p_j d_{ij}$). This formulation assumes that within the set, the citations given by discipline *i* and received by discipline *j* are expected to be proportional to the product of their number of references in the set ($p_{i,j} = p_j p_j$). The observed/expected ratio shows whether the unit under investigation is linking the distant categories within its publication portfolio or not. By using a measure of diversity (Rao-Stirling) in the denominator, we ensure that this measure of coherence is orthogonal to diversity.

*Coherence* $\quad\quad\quad\quad\quad\quad\quad\quad\quad\quad\quad\quad\quad\quad\quad\quad \dfrac{\sum_{i,j} p_{ij} d_{ij}}{\sum_{i,j} p_i p_j d_{ij}}$

*Intermediation*

Intermediation aims to capture the degree to which a given set of publications is distant from the most intensive areas of publication — those dense areas of the map representing the central disciplinary spaces. Since this measure is highly sensitive to artefacts created by the process of classification, we carry out the analysis at a finer level of description, namely the journal level (i.e. we use each journal as a separate category). We propose to use two conventional network analysis measures to characterise the degree to which an organisation's publications lie in these 'interstitial' spaces. The first is the clustering coefficient $cc_i$, which identifies the proportion of observed links between journals over the possible maximum number of links (de Nooy et al., 2005, p. 149). This is then weighted for each journal according to its proportion $p_i$ of publications (or references, or citations), i.e. $\sum_i p_i cc_i$. The second indicator of intermediation is the average similarity of a given journal to all other *N* journals ($\frac{1}{N}\sum_j s_{ij}$) weighted by the distribution of elements ($p_i$) across the categories.[14]

*Average similarity* $\quad\quad\quad\quad\quad\quad\quad\quad\quad\quad\quad\quad\quad\quad \sum_i p_i \left( \dfrac{1}{N} \sum_j s_{ij} \right)$

### 3.3 The assessment of performance

Because the contributions of scientific organisations are so diverse, their evaluation is necessarily complex. It becomes even more so if the evaluator attempts to capture societal contributions (Donovan, 2007; Nightingale and Scott, 2007). Since our research interest lies only in exploring the possible disadvantage that IDR experiences in research assessment (rather than its wider societal impact), we focus on conventional and widely used indicators specifically of scientific performance These measures aim to capture performance according to what Weinberg (1963) called 'internal criteria', i.e. by means of criteria generated within the scientific field.

---

[13] Rafols and Meyer (2010, pp. 273-274) operationalised coherence as the similarity (according to bibliographic coupling) among publications in a set, with the aim of revealing the coherence of topics, rather than disciplinary coherence. Here, we use SCs as units of analysis since the question is whether units are linking or not the *disparate* disciplines in which they publish.

[14] The robustness of the clustering coefficient and the average similarity as indicators of intermediation needs to be confirmed in further studies. They describe low-density landscapes, which are not always associated (as they can be shown to be in this case study) with intermediate or brokering positions. We thank Paul Wouters for this insight.

The first conventional indicator we use is the mean score of the Association of Business Schools' (ABS) journal rankings for the publications of a given research unit. The ABS journal rankings are 'a hybrid, based partly upon peer review, partly upon statistical information relating to citation [i.e. on the Thompson-Reuters Impact Factor], and partly upon editorial judgements' (ABS, 2010, p.1). It has been created by leading academics at BMS belonging to the ABS – thus it follows *internal* criteria. The function of these journal rankings is to indicate 'where best to publish', to inform library purchases and staffing decisions such as 'appointment, promotion and reward committees' and to help to aid 'internal and external reviews of research activity and the evaluation of research outputs' (ABS, 2010, p. 1). In addition to being closely correlated with RAE results, these journal rankings are an explicit part of the BMS 'culture' and are routinely used for recruitment and promotion purposes.

A second conventional indicator is the mean number of citations per publication. Narin and Hamilton (1996, p. 296) argued that bibliometric measures based on citations to publications provide an *internal* measure of the impact of the contribution, and hence a proxy of *scientific* performance. The number of citations per publication (or 'citation impact') is neither an indicator of quality nor importance. Instead, it is a reflection of one form of influence (influence on one's scientific peers) that a publication may exert, which can be used in evaluations provided certain caveats are met (see the detailed discussion in Martin and Irvine, 1983, pp. 67-72; also Leydesdorff, 2008).

One of the key caveats in using citations per paper as a performance indicator is that different research specialties exhibit contrasting publication and referencing norms, leading to highly diverse citation propensities. Hence, some form of normalisation to adjust for such differences between fields is '[p]erhaps the most fundamental challenge facing any evaluation of the impact of an institution's programs or publications' (Narin and Hamilton, 1996, p. 296). The most extensively adopted practice is to normalise by the discipline to which is assigned the journal in which the article is published. Here, the field-normalised figure for citations/paper was calculated by dividing the citations of a given paper by the average citations per paper of the publications of that particular disciplinary category (using data obtained from the 2009 Journal Citation Reports).

For reasons of data availability, we rely on the *Web of Science Subject Categories*[15] as disciplinary categories. Although very unreliable for individual papers, they produce meaningful results for sufficiently large numbers of publications viewed at the scale of global science as a whole (Rafols and Leydesdorff, 2009). One advantage of the Subject Categories is that they are mostly defined at the sub-discipline level (e.g. *Organic Chemistry*), allowing varying degrees of larger 'disciplinarisation' according to their clustering in the global map of science, instead of having to rely on 'essential' disciplinary definitions.

Though widely used, the field normalisation procedure described above is known to be problematic (Leydesdorff and Opthof, 2010). This is, first, because the allocation of journals to disciplines can be made in a number of contrasting but equally plausible ways. There are major discrepancies between various established disciplinary classifications, such as the Web of Science or Scopus categories, which are designed for literature retrieval purposes but are not analytically robust (Rafols and Leydesdorff, 2009). A second reason is because some papers (perhaps especially interdisciplinary ones) may not conform to the conventional citation patterns of a journal; they may, for example, have a 'guest' role in a given category, as in the case of publications on science policy in medical journals. As a result of these difficulties, normalisations using different field delineations (or levels of aggregation) may lead to rather different pictures of citation impact (Zitt et al., 2005; Adams et al., 2008).

---

[15] This study uses the 'Subject Categories' of Web of Science's version 4. Note that in version 5 (as of September 2011), these categories have been relabelled as 'Web of Science Subject Categories', with WC as the new acronym.

To circumvent the problem of delineating the field of a publication, one could instead try to normalise from the perspective of the audience, i.e. via those publications citing the publications to be assessed. One way to normalise from the citing-side is by making a fractional citation count, whereby the weight of each citation is divided by the number of references in the citing publication. Fractional counting was first used for generating co-citation maps by Small and Sweeney (1985). Only recently did Zitt and Small (2008) recover it for the purpose of normalizing for journal 'audience' (following a discussion in Zitt et al., 2005), with Leydesdorff and collaborators subsequently developing this approach for evaluation purposes at the individual paper level (Leydesdorff and Opthof, 2010; Zhou and Leydesdorff, 2011).

Citing-side normalisation can be particularly appropriate for interdisciplinary cases (which receive citations from publications with different citation norms) because it normalises in a way that is not dependent on classifications (Zhou and Leydesdorff, 2011; Zitt, 2011). However, although this corrects for the differences in the number of references in the citing paper, it may not correct for differences in their publication rates. For the purposes of this study, the citing-side normalisation is carried out using only the downloaded citing records (i.e. excluding any citation from the unit being investigated), and then giving each a citation a weight inversely proportional to their number of references, i.e. $\frac{1}{\#\,References}$. Only papers with more than 10 references (including self-citations) are used, since papers with fewer references would have a disproportionately high weight (and in any case these tend not be a 'normal' research publication outlet).

Following conventional practice, in all cases we use the mean to describe the citation distributions. This has long been widely acknowledged to be a flawed method given the highly skewed nature of citation distributions (Narin and Hamilton, 1996, pp. 295-296; Katz, 2000; Leydesdorff & Bornmann, 2011; Leydesdorff and Opthof, 2011). As we will see, it also leads to very high standard errors, which can often render the differences between performance indicators statistically non-significant.

Finally, we also include measures based on the journal Impact Factor, despite wide scepticism of its scientific validity (e.g. Seglen, 1997). We do this for two reasons. First, to examine performance as reflected in a widely used indicator; and second, but more importantly, to check if the results for performance based on using the ABS journal ratings are driven by a substantial reliance on the Impact Factor of journals.[16] We compute the mean Impact Factor of the journals of publications, the mean normalised for the particular Subject Category of the publication journal, and the mean Impact Factor of the citing journals.[17]

---

[16] We thank an anonymous referee for this suggestion.
[17] When a journal is classified into two or more Subject Categories, the average values over these categories are used.

*3.4 Data*

We investigate three centres of IS in the UK: the Manchester Institute of Innovation Research (MIoIR) at the University of Manchester (formerly known as Policy Research in Engineering, Science and Technology, PREST), SPRU (Science Policy Research Unit) at the University of Sussex, and the Institute for the Study of Science, Technology and Innovation (ISSTI) at the University of Edinburgh.

The choice was determined in part by the perceived importance of these centres in the establishment of IS in the UK (Walsh, 2010) and in part by the lack of coverage in the Web of Science of more discursive forms of social science, more reliant on books prevalent in other centres such as the Institute of Science, Innovation and Society (InSIS) at the University of Oxford. These IS units are compared with three leading British BMS: London Business School (LBS), Warwick Business School (WBS) and Imperial College Business School (formerly Tanaka).

The publications of all researchers identified on institutional websites as members of the six units (excluding those holding adjunct, visiting and honorary positions) were downloaded from Thomson-Reuters Web of Science for the period 2006-2010. The downloads were limited to the following document types: 'article', 'letter', 'proceedings paper' and 'review'. Publications by a researcher prior to their recruitment to the unit were also included.[18]

The analysis of all other data relating to journals and Subject Categories is based on the CD-ROM version of the Journal Citation Report, following routines described in previous work (Leydesdorff and Rafols, 2009). A caveat to this approach is that we use the full Web of Science (containing around 11,000 journals) for the units' data, while relying on Journal Citation Reports data (based on approximately 9,000 journals) to carry out parts of the analysis (such as the global maps of science or the field normalisation). In doing this, we are assuming that the structure of the Web of Science and Journal Citation Reports are broadly equivalent.[19]

In order fully to disentangle the analytical results of a unit's publications from the unit's citations, all citing documents from the same unit were removed (i.e. self-citation and citations from institutional colleagues were not included in the citing subset). Due to the retrieval protocol used for the citing papers (which is researcher-based), those papers repeatedly citing the same author were counted only once, whereas those papers citing collaborations between several researchers in the same unit were counted once for each researcher. This inaccuracy only affects the part of the analysis regarding citations (i.e. not the publications or references) and is not expected to result in any serious distortion since intra-organisational collaborations represent only about 10% of publications.

*3.5 Data processing and visualisation*

---

[18] The download was carried out between 20 and 30 October 2010 (except for SPRU publications, which were initially downloaded on 22 May 2010 with an update on 26 October 2010). Additionally, publications citing these researchers' publications were also downloaded in the same period (including those for SPRU).
[19] We thank Thed van Leeuwen at CWTS for making us aware of this potential source of error.

The software Vantage Point[20] was used to process data. A thesaurus of journals to Subject Categories was used to compute the number of aggregated Subject Categories cited in the references (Porter et al., 2007, p. 125). The proportion of references which it was possible to assign in this way ranged from 27% for ISSTI (Edinburgh) to 62% for LBS (London). These proportions are low partly due to variations of journals names among the references that could not be identified, and partly due to the many references to books, lower-status journals and other types of documents not included in the Web of Science. However, the analysis should be statistically robust since between some 1,500 and 10,300 references were assigned to each unit. A minimum threshold of 0.01% of total publications was applied in order to remove Subject Categories with low counts from the variety and disparity measures and thus to reduce the statistical noise. However, no such threshold was applied in calculating the balance, Shannon Entropy and Rao-Stirling measures, since these are computed from proportions, thus are much less affected by small counts.

The ABS ranking for each journal was obtained from the Academic Journal Quality Guide Version 4 (ABS, 2010). The journals are classified into five categories: 1, 2, 3, 4 and 4*. This was used to calculate the average ABS score for each unit. Each level was weighted according to its ascending ordinal position (i.e. 1, 2, 3, 4), while the 4* category was given a weight of 5. In addition, Subject Categories were assigned to all journals in the ABS lists that are indexed in the Journal Citation Reports, these amounting to 60% of those on the ABS list. These data were used to produce overlay maps with the distributions of journals over Subject Categories corresponding to each ABS category. The number of citations/paper was computed using the field *Times Cited* (TC) in the Web of Science record.[21] Intermediation measures were computed with Pajek using the journal similarities matrix. The average clustering coefficient (for a two edges neighbourhood, i.e. $CC_2$ routine in *Pajek*) was computed using a threshold value of 0.2.

The freeware Pajek[22] (de Nooy et al., 2005) was used to construct all the networks except those in Figure 5. First, disciplinary overlay maps were produced by setting the size of each node proportional to the number of references in a given Subject Category, as explained in Rafols et al. (2010)[23], using 2009 data for the base-map (grey background). Second, cross-citations maps (green links) between Subject Categories were generated and overlaid on the disciplinary maps in order to generate Figure 3. Lines are only shown if they represent a minimum of 0.2% of citations and more than five times (these were s ad-hoc choices based on trial and error) the expected proportion of cross-citations among Subject Categories in comparison to average Web of Science cross-citation flows. This shows the extent to which the relations between disciplines are novel or, on the contrary, already well established.

---

[20] http://www.thevantagepoint.com
[21] As a result of the earlier download of SPRU data in May 2011, the *Times Cited* field of SPRU papers had to be extrapolated. The extrapolation was carried out as follows. In October 2010, 730 unique papers citing SPRU (Sussex) papers were found in the Web of Science. For the other five units there was an average discrepancy of 8.5% between the unique papers found in Web of Science citing them, and the counts in TC (the TC being larger because each unique citing paper can reference various publications of the same unit). By using this average discrepancy, 792 citations (730 citations plus the 8.5% discrepancy) were estimated for SPRU. The possible inaccuracy introduced by this extrapolation is well within the standard error (~10%).
[22] http://pajek.imfm.si
[23] Details of the method are available at http://www.leydesdorff.net/overlaytoolkit/. After submission of this article, Leydesdorff and Rafols (2012) developed a new application that produces overlays based on journal-based global maps of science instead of using Subject Categories. These are available at http://www.leydesdorff.net/journalmaps/.

The freeware VOSviewer[24] (Van Eck and Waltman, 2010) was used to produce a journal map in a journal density format (a map in which red areas represents a local high density of journals, and blue low density). A sub-set of 391 journals was constructed from the journals in which each unit published (excluding journals that contributed less than 0.5% of the publications for each unit) and the top 100 journals which all units (collectively) cited. The cross-citations between these journals were obtained from the 2009 Journal Citation Reports. This was used to compute the cosine similarities matrix in the cited dimension, which was then inputted into VOSViewer (for details see Leydesdorff and Rafols, 2012). The size of nodes was determined by the number of publications (or references or citations) per journal, normalised according to the sum of all publications (or references or citations), and overlaid on the base-map.

## 4. Results

### 4.1 Interdisciplinarity of organisational units

The following sections present the results of this investigation. First we show that IS units are more interdisciplinary than BMS according to three different perspectives and their associated metrics.

*Diversity and coherence*

Figure 3 shows the overlay of the publications of ISSTI (Edinburgh, top) and LBS (London, bottom) over the global map of science – as a representative illustration of the findings in this analysis regarding the general contrast between the three IS units (including ISSTI) and the three comparator BMS (including LBS). The full set of diversity maps for all the organisations can be found at www.interdisciplinaryscience.net/maps and in the supplementary materials.[25] The set of overlay maps were generated for each of the six units and then for the Subject Categories of publications, references and citations (excluding self-citations). These results show that IS units are cognitively more diverse in the sense that they spread their publications (along with the associated references and citations) over a wider set of disciplines (i.e. there is greater 'variety'), do so more evenly (i.e. exhibit greater 'balance') and across larger cognitive distances (i.e. show more 'disparity'). No significant time trends were found. The differences are more pronounced in the case of publications and citations than for references,[26] which tend to be relatively widely spread for both IS and BMS. These insights are shown in the form of indicators in Table 1 and Figure 4.

---

[24] http://www.vosviewer.com
[25] http://www.sussex.ac.uk/Users/ir28/IDR/Disciplinary_Diversity.pptx
http://www.sussex.ac.uk/Users/ir28/IDR/Disciplinary_Coherence.pptx
[26] In the case of IS publications and references, one might speculate that the higher diversity observed is just the circumstantial result of their involvement in a field, IS, where the subject of research happens to be another science, and hence is citable. The results for the diversity measures of citing articles (after excluding self-citations within units) are important because they would seem to refute the possibility that the larger diversity of IS publications is caused solely by references to the disciplines of the subject matter of the article (e.g. health or energy) rather than genuine scholarly engagement with more distant disciplines. We further validated this view by examining abstracts of articles citing IS units from the natural sciences or engineering. The sample revealed that these articles included both conventional publications embedded in the discipline and policy or opinion papers reflecting on the topics (which might be by other IS scholars). Further research is needed to understand the role of publications and references by IS scholars in the context of 'practitioner' journals.

Second, not only are IS units more diverse, but their publications cite more widely across distant Subject Categories than BMS. This is shown by the green links overlaid in Figure 3 (representing cross citations between Subject Categories more than five times the expected in the global map of science). ISSTI (Edinburgh) has major citation flows between management and biomedical sciences, which are rare in the global citation patterns. SPRU (Sussex) between economics and planning, on the one hand, and ecology, environment and energy, on the other. This is evidence that these IS units are not only diverse in the sense of 'hosting' disparate disciplines, but are also directly linking them. In particular, they play a bridging role between the natural sciences and social sciences.

By contrast, the leading BMS examined here are not only less diverse, but also more fragmented (or less coherent) in disciplinary terms, in the sense that they tend to cite more within specialties or disciplines. For example, Imperial is the most diverse of the BMS, thanks in part to its research groups on IS and healthcare management. However, this latter line of research is not strongly linked to other social sciences at Imperial, as shown by the relative scarcity of cross-citations. In this case, then, co-location of health research and management in the same BMS does not appear to lead to interdisciplinary exchange. The bridging function between the natural sciences and social sciences carried out by IS units is captured by the coherence indicator shown in Table 1 and illustrated in Figure 4.

Measures such as diversity may exhibit size effects, i.e. they may increase or decrease depending of the size of department. Since the IS units are between two to four times smaller than BMS, one might wonder whether size-effects may explain the differences in the diversity measures. However, the most obvious size effect one might expect would be for larger units to display greater diversity, given the higher probability of having a very small proportion of publications/references/citations in some Subject Categories. Since the observed relation is the inverse, i.e. the smaller units exhibit the highest diversity, one can be confident that the results are not an indirect effect of size. Indeed, they are evident despite such an effect, and are likely to be correspondingly stronger if size were controlled for. (There is no size effect expected in the case of coherence, given that it is computed from a ratio.)

**ISSTI**

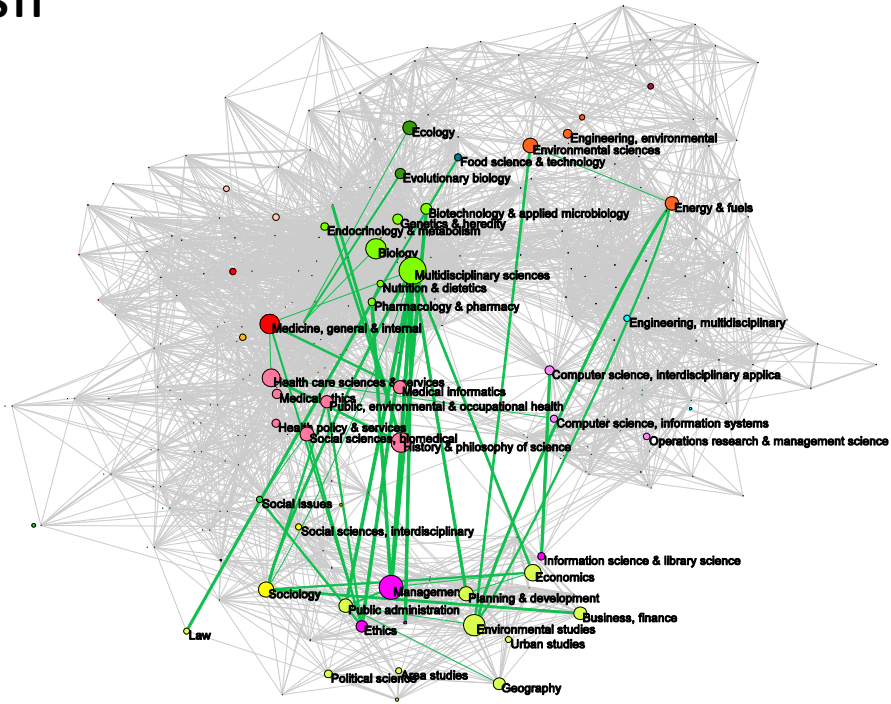

**LBS**

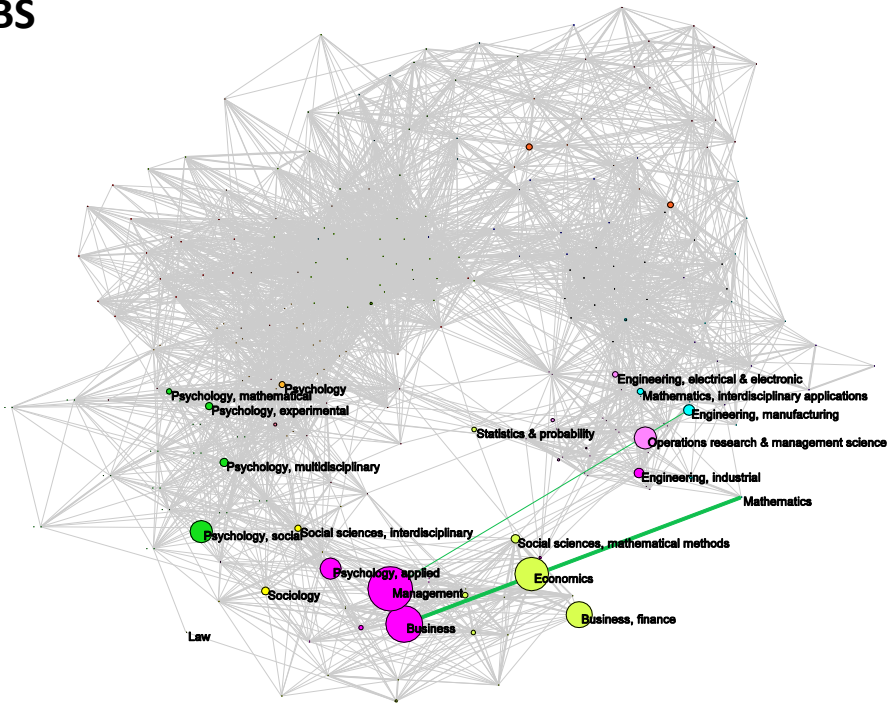

**Figure 3. Overlay of number of references on Subject Categories (source) by of ISSTI (Edinburgh, top) and LBS (London, bottom) on the global map of science.** The extent of referencing (or citing) between Subject Categories (as indicated by green links) by a given unit is shown only for observed values five times larger than expected**.** Each node represents a sub-discipline (Subject Category). Grey lines indicate a certain level of similarity between Subject Categories. The degree of superposition in the grey background illustrates the degree of similarity between different areas of science for all 2009 Web of Science data. Diversity of references (as reflected in the spread of nodes over map) and referencing across disparate Subject Categories (the amount of cross-linking) are interpreted as signs of interdisciplinarity.

**Table 1. Indicators of diversity and coherence for each organisational unit**

|  | Innovation Studies (IS) Units | | | Business and Management Schools (BMS) | | |
|---|---|---|---|---|---|---|
|  | ISSTI Edinburgh | SPRU Sussex | MIoIR Manchester | Imperial College | WBS Warwick | LBS London |
| # of Publications | 129 | 155 | 115 | 244 | 450 | 348 |
| **Diversity of Subject Categ. of Publications** | | | | | | |
| Variety | 28 | 20 | 19 | 15 | 20 | 9 |
| Balance | 0.89 | 0.83 | 0.82 | 0.74 | 0.74 | 0.72 |
| Disparity | 0.83 | 0.84 | 0.82 | 0.79 | 0.77 | 0.77 |
| Shannon Entropy | 3.56 | 3.24 | 2.97 | 2.97 | 3.08 | 2.34 |
| **Rao-Stirling Diversity** | **0.81** | **0.78** | **0.73** | **0.72** | **0.68** | **0.60** |
| # of References | 1737 | 2409 | 1558 | 6017 | 8044 | 10381 |
| **Diversity of Subject Categ. of References** | | | | | | |
| Variety | 28 | 18 | 17 | 17 | 20 | 15 |
| Balance | 0.82 | 0.71 | 0.70 | 0.65 | 0.62 | 0.57 |
| Disparity | 0.83 | 0.84 | 0.85 | 0.83 | 0.78 | 0.83 |
| Shannon Entropy | 4.12 | 3.58 | 3.38 | 3.25 | 3.15 | 2.80 |
| **Rao-Stirling Diversity** | **0.83** | **0.79** | **0.73** | **0.73** | **0.69** | **0.68** |
| # of Citations | 316 | 767 | 419 | 1229 | 1246 | 1593 |
| **Diversity of Subject Categ. of Citations** | | | | | | |
| Variety | 32 | 21 | 22 | 20 | 24 | 15 |
| Balance | 0.90 | 0.78 | 0.78 | 0.73 | 0.73 | 0.65 |
| Disparity | 0.85 | 0.84 | 0.84 | 0.82 | 0.80 | 0.77 |
| Shannon Entropy | 4.22 | 3.72 | 3.42 | 3.48 | 3.50 | 2.99 |
| **Rao-Stirling Diversity** | **0.85** | **0.81** | **0.77** | **0.76** | **0.74** | **0.68** |
| **Citations between Subject Categories** | | | | | | |
| **Coherence** | **0.73** | **0.75** | **0.80** | **0.60** | **0.66** | **0.54** |

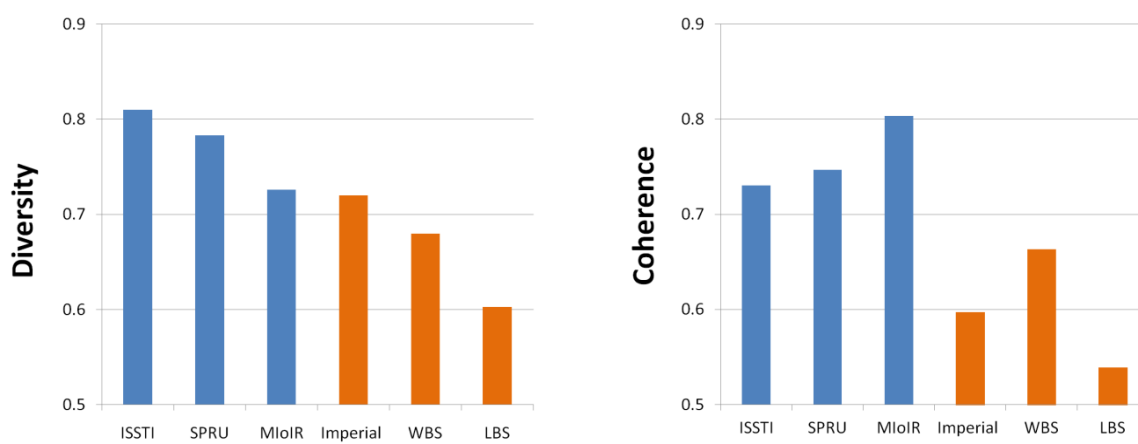

**Figure 4. Indicators of diversity (Rao-Stirling) and coherence for the publications by organisational unit**

*Intermediation*

The third property of IDR that we want to investigate is whether a given body of research lies within or between existing disciplinary boundaries. For this purpose the Web of Science Subject Categories are too coarse. Instead of using the Subject Category disciplinary maps, we created maps of the main 391 journals in which the six units examined here publish and reference (see the methods section). We used the density visualisation option of the software VOSviewer, which is helpful in distinguishing between the dense areas associated with disciplinary cores (depicted in red) and sparser interstitial areas associated with IDR (in green). To produce the base-map, Subject Category to Subject Category cross-citation data from the 2009 Journal Citation Reports were used to generate a similarity matrix, which then served as an input for the visualisation programme. The publications, references and citations associated with each unit were then overlaid on this map. Note that this map is constructed on a different basis from conventional journal maps (where the relative positions reflect direct similarities within the local data rather than the position of the local map in a similarity space created using all the Web of Science).

The local journal maps of IS-BMS (see Figure 5 and website [27]) show three poles: management, economics, and natural sciences. The sharp polarisation between economics and management is fully consistent with the findings by Van Eck and Waltman (2010, pp. 529-530).[28] Interestingly, *Research Policy*, which was identified as the most important and central journal of IS by Fagerberg et al. (this issue), occupies an equidistant position between the management and the economics poles – and slightly tilted towards the natural sciences.

The third pole encompasses the various specific natural sciences studied by these units. The map reveals that, within the combined IS-BMS context, journals of different natural sciences are cited similarly, in comparison to the differences among the citations to social science journals. Thus, unlike the economics and management areas, this third pole can probably be interpreted as an artefact generated by the local perspective (i.e. a too small subset of natural science journals) rather than a genuine disciplinary core in its own right. This pole is nevertheless useful since it provides a means to show the degree of interaction between the social sciences and the natural sciences. Journals that are more oriented to science rather than innovation, such as *Social Studies of Science* and *Scientometrics*, are closer to this pole. Overall, the relative position of the different disciplines in Figure 5 is quite consistent with that of the global map of science seen in Figure 3, but here some areas such as business and economics have been 'expanded', while the natural sciences have been compressed. The effects of these shifting spatial projections are neutral with respect to the conclusions drawn here.

The overlay maps in Figure 5 show that BMS units mainly publish, reference and are cited by journals in the dense areas of management and economics. IS units, in contrast, have most of their activity in the interstitial areas lying between management, economics and the natural sciences, that is, in journals such as *Research Policy*, or in journals of application areas such as *Social Science and Medicine* or *Energy Policy*. These differences across units in terms of the position of the journals in which they publish can be expressed quantitatively by means of the indicators 'Clustering coefficient' and 'Average similarity' (defined in Section 3.2) of the journals as described in Table 2 and Figure 6. In summary, what the journal maps show is that

---

[27] http://www.sussex.ac.uk/Users/ir28/IDR/Intermediation.pptx
[28] See Van Eck and Waltman's (2010) interactive maps at
http://www.vosviewer.com/maps/economics_journals/

IS units perform their boundary-spanning role, at least in part, through interdisciplinary journals.

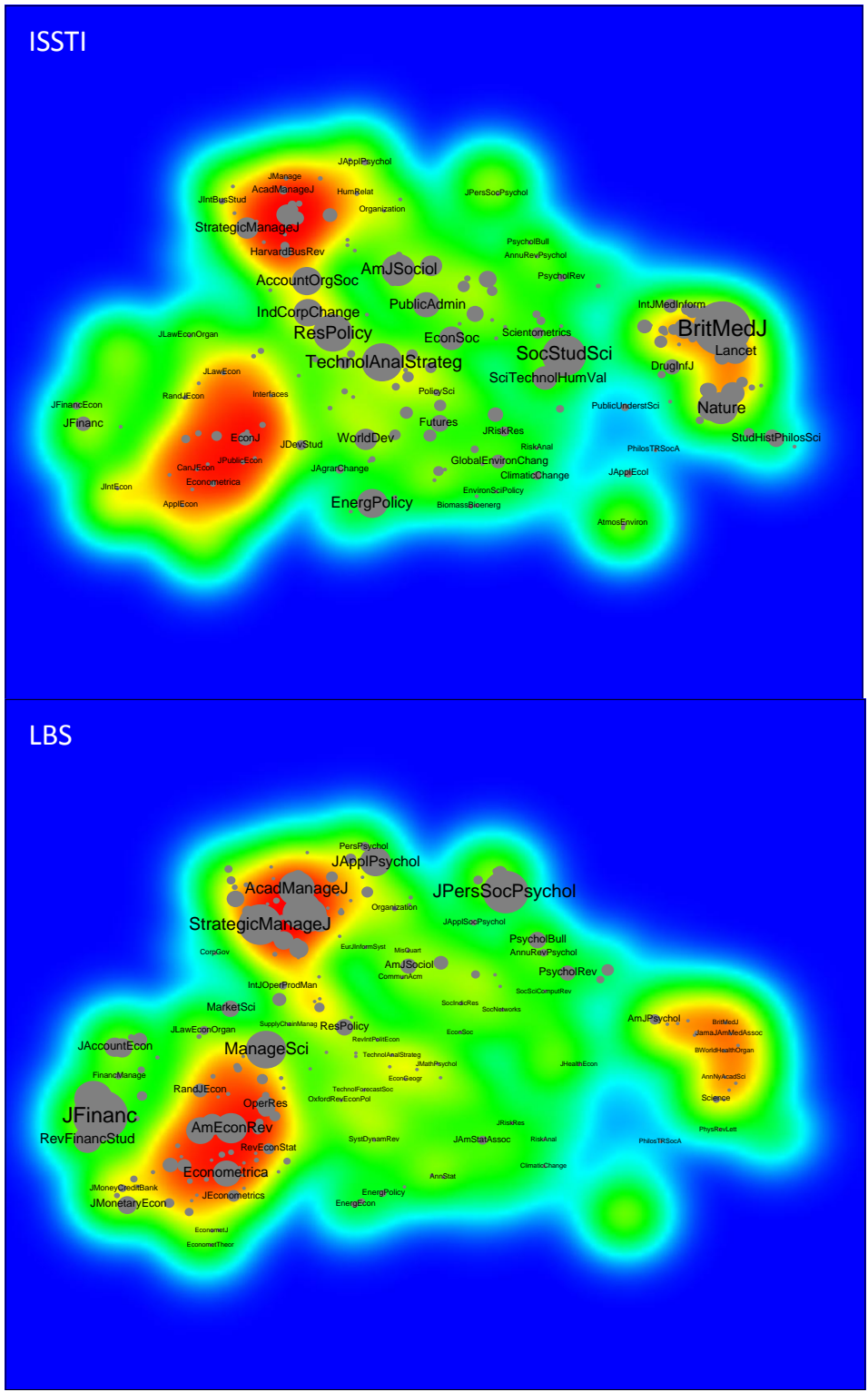

**Figure 5. Overlay of the references of ISSTI (Edinburgh) and LBS (London) publications in a local journal map.** The map illustrates the similarity structure of the 391 most important journals for all six IS and BMS units analysed. Red areas correspond to a high density of journals, indicating areas of mono-disciplinary activity. Green areas show low density. Node size indicates the proportion of a unit's references in a given journal. Journals located in between red areas, i.e. between disciplinary cores, are interpreted as interdisciplinary. (This figure needs to be viewed in colour).

**Table 2. Indicators of intermediation by organisational unit**

|  | Innovation Studies (IS) Units | | | Business and Management Schools (BMS) | | |
| --- | --- | --- | --- | --- | --- | --- |
|  | ISSTI Edinburgh | SPRU Sussex | MIoIR Manchester | Imperial College | WBS Warwick | LBS London |
| **Journals of pub's** | | | | | | |
| **Clustering coefficient** | 0.128 | 0.098 | 0.075 | 0.189 | 0.165 | 0.202 |
| **Average similarity** | 0.028 | 0.034 | 0.036 | 0.050 | 0.045 | 0.060 |
| **Journals of references** | | | | | | |
| **Clustering coefficient** | 0.178 | 0.182 | 0.166 | 0.236 | 0.221 | 0.235 |
| **Average similarity** | 0.044 | 0.050 | 0.058 | 0.066 | 0.065 | 0.068 |
| **Journals of citations** | | | | | | |
| **Clustering coefficient** | 0.120 | 0.096 | 0.074 | 0.157 | 0.167 | 0.183 |
| **Average similarity** | 0.029 | 0.034 | 0.037 | 0.046 | 0.044 | 0.055 |

Note: low values for each metric indicate higher levels of intermediation. Standard errors are not provided because they are negligible (all smaller than 0.07%).

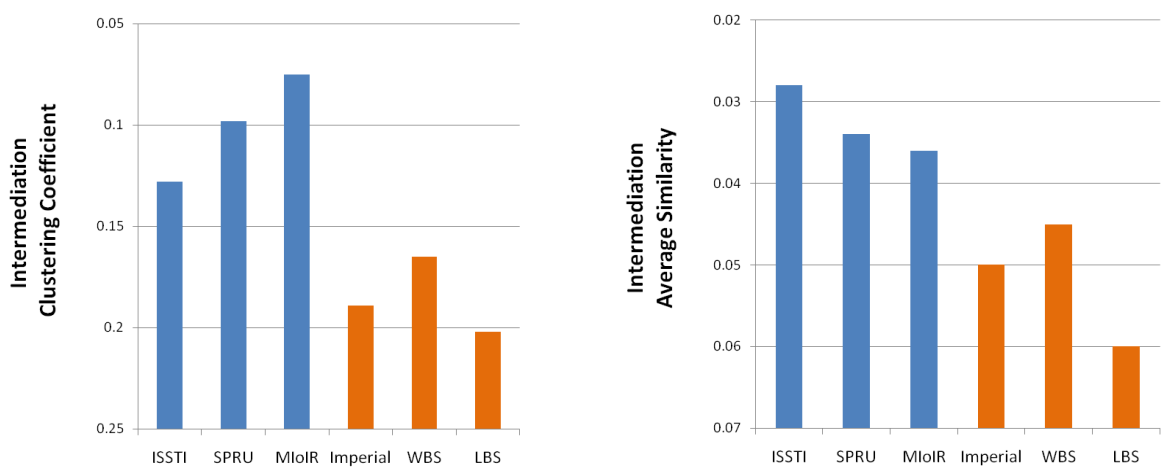

**Figure 6. Indicators of intermediation of publications by organisational unit**

*4.2 Disciplinary bias in the ABS journal rankings*

Now we turn our attention to the disciplinary profiles of the journals under different rating categories in the ABS classification. For each rank, from 1 (the lowest quality), to 4* (the highest), we use the Journal Citation Reports to assign journals to Subject Categories. The Journal Citation Reports coverage of the ABS journals was low for rank 1 (14%), but reached an acceptable level for rank 2 (56%), and was almost complete at the highest ranks. We analyze the disciplinary diversity of each rank in terms of its distribution of journals in Subject Categories, following the same measures (section 3.2) and data protocol (section 3.5) as for the analysis of organisational units, only now the unit of analysis is journals rather than articles. The results are shown in Table 3 and Figures 7 and 8 (full map set is available in website[29]).

---

[29] http://www.sussex.ac.uk/Users/ir28/IDR/ABS_Ranking_Diversity.pptx

Acceptable Standard (Rank 2)

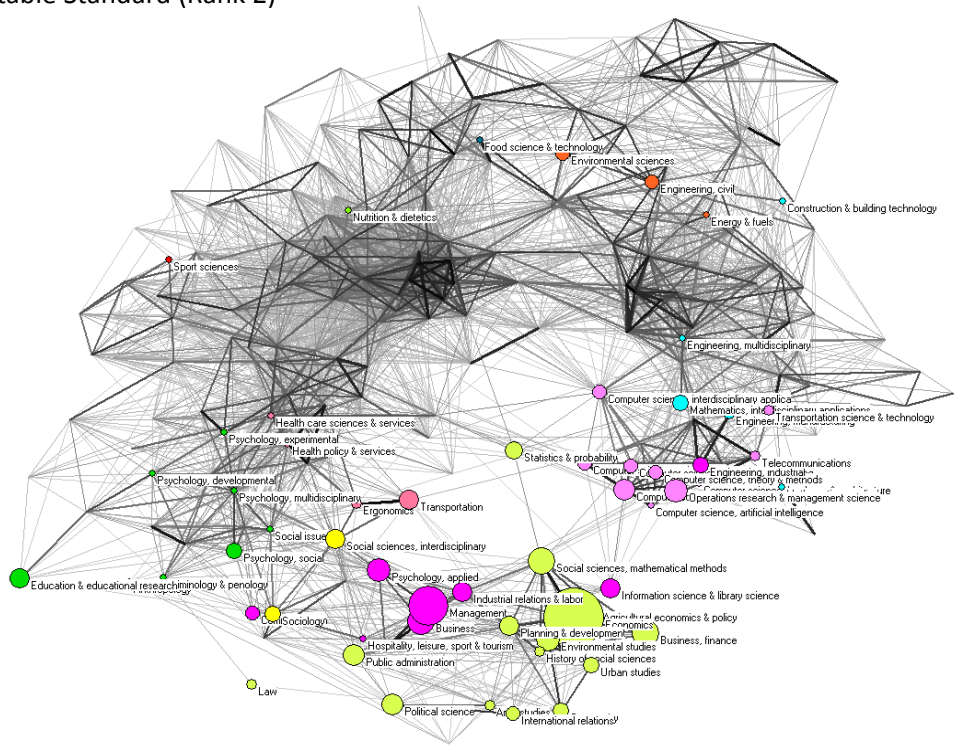

World Elite (Rank 4*)

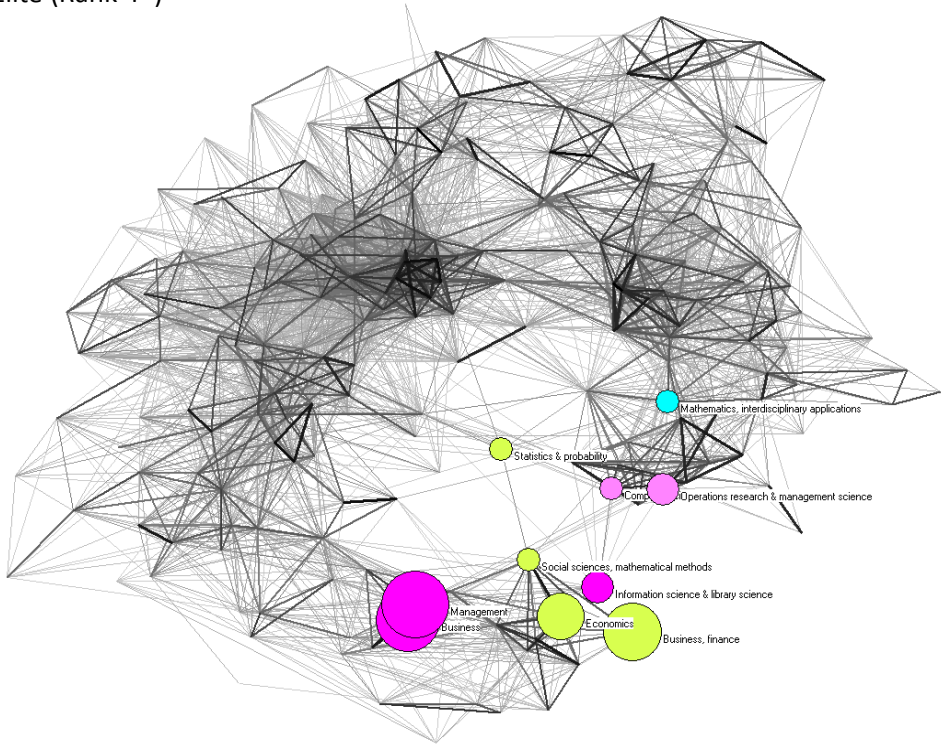

**Figure 7. Distribution of journals across different categories for Association of Business Schools (ABS)' Rank 2 ('Acceptable Standard') and Rank 4\* ('World Elite')**

**Table 3. Disciplinary diversity indicators of the Association of Business Schools (ABS) ranks**

|  | Rank 1 'Modest standard' | Rank 2 'Acceptable standard' | Rank 3 'Highly regarded' | Rank 4 'Top in Field' | Rank 4* 'World Elite' |
|---|---|---|---|---|---|
| **# of Journals in Journal Citation Reports** | 29 | 166 | 199 | 73 | 21 |
| **Diversity of Subject Categories of Journals** | | | | | |
| Variety | 27 | 58 | 56 | 31 | 10 |
| Balance | 0.90 | 0.85 | 0.81 | 0.86 | 0.87 |
| Disparity | 0.87 | 0.82 | 0.79 | 0.79 | 0.77 |
| Shannon Entropy | 2.98 | 3.45 | 3.28 | 2.94 | 2.00 |
| **Rao-Stirling Diversity** | **0.78** | **0.73** | **0.70** | **0.69** | **0.57** |

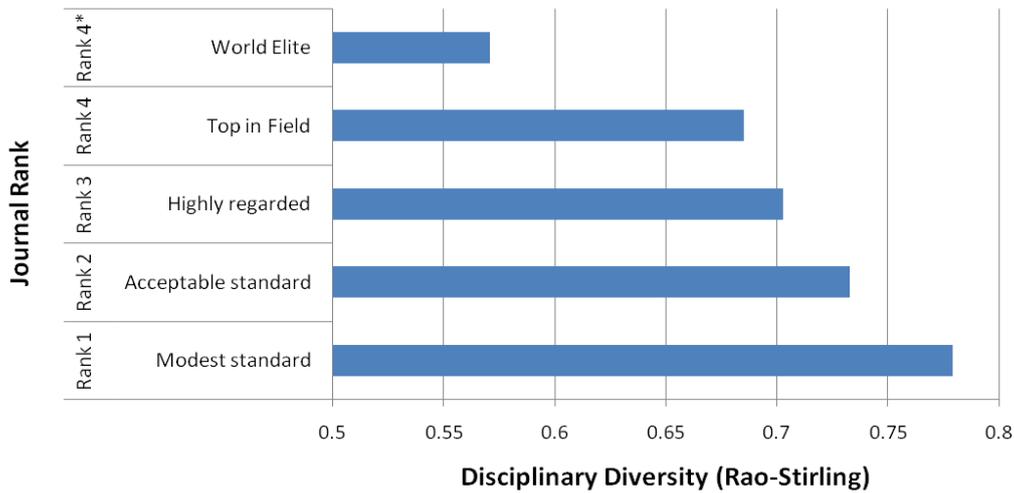

**Figure 8. Diversity of the disciplinary distribution of journals for each rank of the Association of Business Schools (ABS)**

These data show that the highest ranking journals are much less diverse than the lowest ranking ones. In particular, the top rank (4*) narrowly focuses on three Subject Categories: management, business and finance. Lower ranks are spread across various social sciences, including economics, geography, sociology, psychology, and some engineering-related fields such as operations research and information science, as well as some applications such as environment or food. Thus, while the ABS list includes journals from many disciplines, only some of those in their core subject matters are perceived by ABS as 'World Elite' journals.[30]

---

[30] The fact that ranks 4* and 4 only contain 21 and 73 journals respectively, in comparison to more than 100 journals in ranks 3 and 2, might partly explain why the higher ranks are less diverse. This certainly has an effect on the number of SCs and some effect on the Rao-Stirling and Shannon diversity. However, the key insight here comes from understanding the extent to which the various Subject Categories of the top-rank journals are associated with the same broad disciplines or not. Figure 7 suggests that highly ranked journals in the ABS list cover a smaller region in the science maps. This should have been clearly reflected in the measure of disparity. However, the differences observed in disparity between ranks are minor. This lack of clear differentiation indicates that the distance metric we use is only sensitive to short-range differences between Subject Categories

*4.3 Performance of organisational units*

Finally, we can now explore how the disciplinary bias in the ABS journal rankings affects the assessment of organisational units. To do this, we calculated the mean of the scores of the journals in which the units publish. In doing so, we first face a problem of assignation: whereas only 43% of ISSTI (Edinburgh) or 51% of SPRU (Sussex) journals that are listed in the Web of Science are also included in the ABS list, the coverage reaches 79% and 93% of Web of Science journals in the case of WBS (Warwick) and LBS (London), respectively. The results are shown in Table 4 and Figure 9 (see also website[31]). They conclusively show that the three BMS perform significantly better than the IS units. Within the BMS, the narrow disciplinary profile of LBS achieves a much higher figure than the other two BMS. This is associated with the strong negative Pearson correlation between degree of interdisciplinarity across any metrics and ABS-based performance: -0.78 (Rao-Stirling diversity), -0.88 (coherence), 0.92 (Intermediation, clustering coefficient).

**Table 4. Performance indicators**

|  | Innovation Studies (IS) Units | | | Business and Management Schools (BMS) | | |
|---|---|---|---|---|---|---|
|  | ISSTI Edinburgh | SPRU Sussex | MIoIR Manchester | Imperial College | WBS Warwick | LBS London |
| **ABS journal ranking-based** Mean (Standard Error) | | | | | | |
| Mean ABS rank | 2.82 (0.13) | 2.65 (0.10) | 2.54 (0.10) | 3.36 (0.07) | 3.01 (0.05) | 3.92 (0.05) |
| % Papers ranked | 43% | 51% | 74% | 69% | 79% | 93% |
| **Citation-based** Mean (Standard Error) | | | | | | |
| Citations/paper | 2.69 (0.45) | 5.11 (0.59) | 3.50 (0.63) | 5.30 (0.73) | 2.91 (0.23) | 5.04 (0.39) |
| Citations/paper (Journal normalized) | 1.99 (0.31) | 2.74 (0.36) | 2.35 (0.34) | 2.69 (0.33) | 2.16 (0.16) | 2.28 (0.17) |
| Citations/paper (Field normalized) | 1.67 (0.28) | 2.79 (0.35) | 2.10 (0.43) | 3.34 (0.47) | 2.11 (0.16) | 3.60 (0.28) |
| Citations/paper (Citing-side normalized) | 0.18 (n.a.) | 0.12 (n.a.) | 0.09 (n.a.) | 0.13 (n.a.) | 0.07 (n.a.) | 0.11 (n.a.) |
| **Impact Factor-based** Mean (Standard Error) | | | | | | |
| Journal Impact Factor | 2.29 (0.38) | 3.14 (0.51) | 1.96 (0.34) | 2.76 (0.27) | 1.65 (0.09) | 2.50 (0.09) |
| Journal Impact Factor (Field normalized) | 1.17 (0.12) | 1.26 (0.11) | 0.98 (0.06) | 1.46 (0.07) | 1.11 (0.03) | 1.74 (0.06) |
| Citing journal Impact Factor | 3.12 (0.28) | 2.45 (0.15) | 1.98 (0.11) | 2.79 (0.14) | 1.79 (0.06) | 2.18 (0.05) |

Note: The standard deviations for the figures on citations/paper using citing-side normalisation are not available because the data on the citing articles were collected in an aggregate form.

Next we compare the ABS-based performance with citation-based performance. We should emphasize that this analysis is only exploratory. Since we are counting citations received by groups of papers published during the period from January 2006 to October 2010 and

---

(i.e. it gives similar large distances when measuring Business to Economics, and Business to Astronomy). This suggests that there is scope for improving the distance metric used (Leydesdorff and Rafols, 2011a).
[31] http://www.sussex.ac.uk/Users/ir28/IDR/Performance_Comparison.pptx

analysing the citations they received up until October 2010 instead of using a fixed 'citation window', the results should be interpreted as only indicative.[32] Although imperfect, the estimates obtained should be sufficiently robust to provide tentative insights and illustrate the inherent difficulties and ambiguities of using citation-based performance indicators.

First, it is important to notice that the standard error is extremely high (in the range of 8-18%) – so high that ranking the units becomes problematic. This is the consequence of the conventional statistical (mal)practice of using the mean to describe skewed distributions (Leydesdorff and Bornmann, 2011). Given these high statistical deviations, even for large schools, it is somewhat surprising that citation-based research rankings are used so prominently by BMS and by the *Times Higher Education* when ranking universities or departments, without reporting the degree (or lack) of statistical significance.

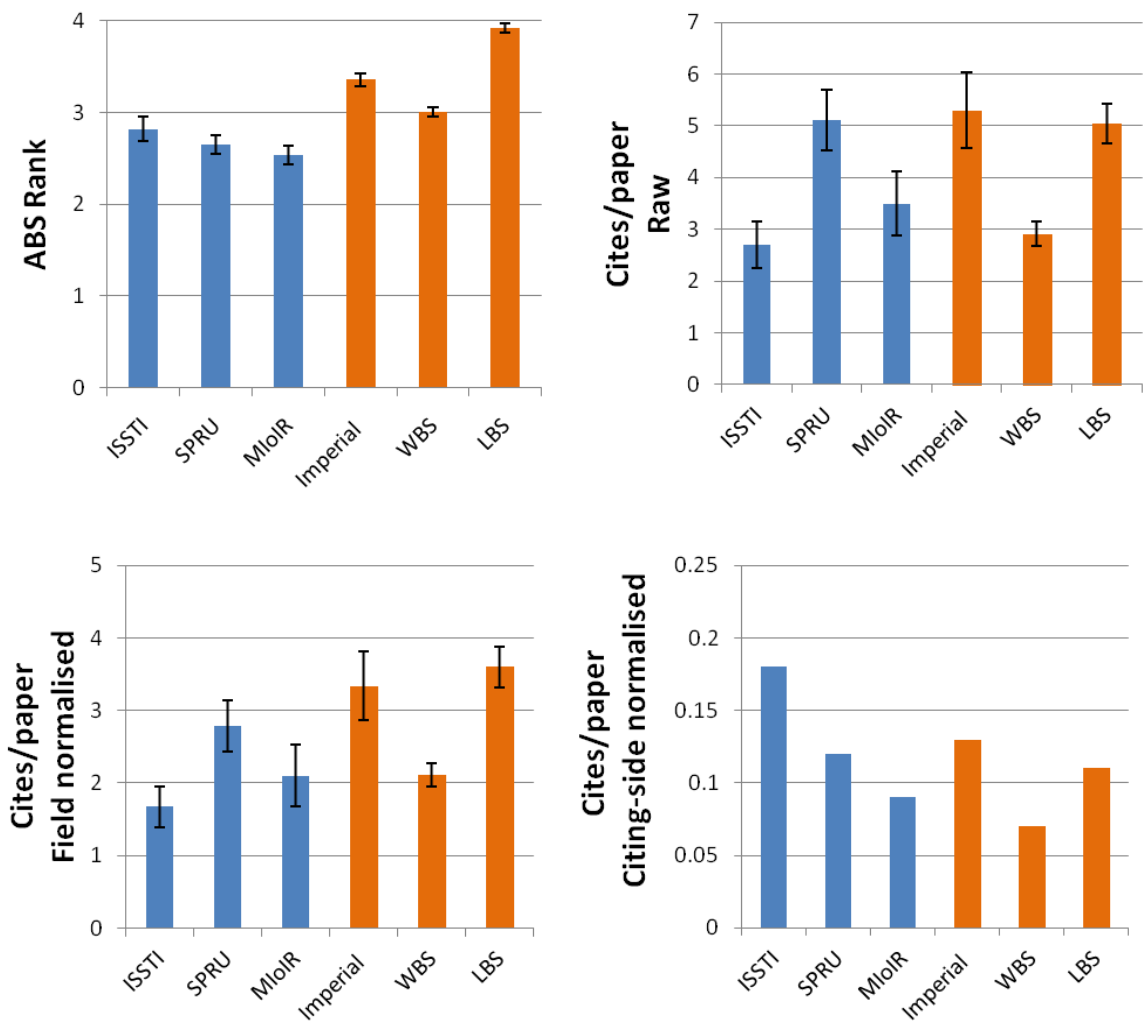

**Figure 9. Performance indicators**

---

[32] Using a fixed 'citation window' means studying the citations that each paper received for a fixed number of years after its publication. The disadvantage of this method is that it only allows studies of past research. In this case, we should have studied publications produced over the period 2001-2005 in order to allow for a 5-year citation window for each document. But doing so would have resulted in an outdated portrayal of the units' performance, as well as encountering major hurdles in the data-gathering due to researchers changing jobs.

The analysis shows, first, that BMS units do not perform better than IS units in terms of total number of citations. Second, normalisation by journal Impact Factor shows that IS units and BMS have similar citation frequencies within the journals in which they publish (shown in website[33]). Third, a field-based normalisation slightly lowers the performance of IS units in comparison to untreated (*raw*) counts. One can put forward a possible explanation for this result: if IS papers are normalised by the field in which they publish, they are doubly disadvantaged in respect both of their publishing in natural sciences (because even if they receive many citations – all else being equal – they tend to be less cited than natural science papers in those journals), or in the social sciences (because they face disproportionate difficulties in publishing in the most prestigious journals – i.e. those which tend to accrue more citations).[34] Fourth, we use the fractional-counting for citation normalisation, which proportionately reduces the value of each citation by the number of references in the citing paper. With this form of normalisation, the correlation between citation-based and ABS-based performance completely vanishes. We highlight this as an important policy-relevant result.

Finally, following the tenet of using multiple indicators where possible, we also estimate the performance based on journal Impact Factor values. We should stress that Impact Factor-based measures have been convincingly shown to be a worse indicator of quality than citations (Seglen, 1997). Nevertheless, in this case the findings are quite similar to those obtained from citations. Overall, the mean Impact Factors of IS publications and citing papers are as high as those of BMS. Interestingly, the standard error of Impact Factors is much higher in IS units than in BMS, which is indicative of more diverse publication practices. Again, upon normalisation by field (Subject Category) of publication, the relative performance of IS units is somewhat reduced, while under the citing field perspective their performance remains comparatively strong.

In summary, based on the ABS journal ratings, all three BMS units show significantly better performance than the IS units. However, a re-examination of this using other conventional bibliometric measures does not provide such a clear result. Raw citation and Impact Factor measures place SPRU (Sussex) at about the same level as Imperial and LBS (London), and MIoIR (Manchester) and ISSTI (Edinburgh) on a par with WBS (Warwick). The citing-side normalisation completely reverses the results, pointing to ISSTI as the best performer. Comparisons based on field normalisation place Imperial and LBS slightly ahead, but without a statistically significant lead over SPRU given the high standard errors. In short, these results show how different, but prima facie equally legitimate, metrics can yield fundamentally different conclusions.[35]

One may speculate as to what might account for the relative drop in the performance of IS units when judged by ABS journal-rankings. On the one hand, there appears to be a bias in the ABS list associated with the focus on certain dominant Business & Management journals – as described in the previous subsection. On the other hand, there may be a more general mechanism, reflecting the greater difficulty that interdisciplinary papers face in being accepted in mainstream disciplinary fields compared to disciplinary papers of the same quality. Although it needs further confirmation, this explanation is apparently supported by the high correlation observed (at the aggregated unit level) between the ABS-based ranks of the six organisations and their field-normalised performance, either in terms of the field-normalised Impact Factor data (0.920, p=0.009), or (although much less significant) for field-normalised data on citations/paper (0.765, σ=0.076; compared with a higher correlation of 0.922 and σ=0.009 between with the untreated figures for citations/paper and field-normalised citations/paper).

---

[33] http://www.sussex.ac.uk/Users/ir28/IDR/Performance_Comparison.pptx

[34] This explanation is supported by the observation that IS publications have less citations for high Impact Factor journals (1.39 times the Impact Factor of journals in comparison to 1.76 for BMS, if the journal Impact Factor is larger than 5) yet more citations from low Impact Factor journals (4.20 times the Impact Factor of journals in comparison to 2.96 for BMS, if the Impact Factor is lower than 0.5).

[35] It is worth noting that the results are much more stable if we look only at the relative performance of BMS compared with one another. In this case, LBS (London) and Imperial obtain similar results, with WBS (Warwick) coming third. This observation supports the interpretation that the contrasting results for IS units are due to differences in disciplinary make-up.

These findings also have implications for bibliometric performance measures. Although still somewhat of an open issue, more sophisticated studies seem to suggest that citing-side normalisation provides a more robust measure of citation impact, since it offers a more accurate description of the citation context of each individual paper (Zitt and Small, 2008; Zhou and Leydesdorff, 2010). The differences in results we find in this study suggest that further research is needed to investigate whether the conventional field normalisation has been systematically under-estimating interdisciplinary contributions in comparison with the results obtained with this newer normalisation method.

The picture that emerges from all the indicators would seem to support the call for more rigorous 'plural and conditional' forms of appraisal mentioned earlier (Stirling, 2008; 2010) that directly address the need to employ 'converging partial indicators' in research evaluation (Martin, 1996; Martin and Irvine, 1983). If rankings are determined more by the choice of indicator than by the content of the research, with those indicators being open to intentional or unintentional bias, then the objectivity of such rankings will remain questionable. Similar warnings about the inconsistencies between performance indicators and how they depend on the size of field or the classification methods used in the citation-normalisation procedure have been repeatedly voiced in the past (Adams et al., 2008; Leydesdorff, 2008; Zitt, 2005; Zhou and Leydesdorff, 2011). This paper confirms just how problematic the use of a single metric can be.

Yet despite such differences, certain conclusions can nevertheless be drawn from this study. First, the IS units are clearly more interdisciplinary than the BMS considered here. Second, the performance of IS units is significantly undervalued in the ABS metrics when compared to a range of citation and Impact Factor metrics. While the ABS measure of strong performance is seemingly associated with a narrower disciplinary focus (on business, management, finance and economics), it is not necessarily related to a stronger performance in terms of citations. This is of some concern, since citations are generally considered to be a more reliable performance indicator than journal-based measures (Seglen, 1997).

## 5. Discussion: How the bias in rankings can suppress interdisciplinary research

### 5. 1 Mechanisms of bias amplification

Although the forthcoming UK research assessment exercise (RAE, now retitled the Research Excellence Framework) does not officially rely on journal rankings, the widespread perception, at least in the field of Business & Management, is that the number of publications in top journals (as judged by ABS in this case) will strongly influence the outcome. As noted previously, various studies have shown this was the case for the 2008 assessment (Taylor, 2011, pp. 212-14; Kelly et al., 2009; David Storey, personal communication, March 2011).[36]

A number of complementary distorting mechanisms may further amplify the bias against IDR apparent in these results. The first is that the percentage of publications appearing in ABS-listed journals is much lower for IS units than for BMS (see Figure 10). If each researcher is expected to submit four articles, then the *average* researcher in an IS unit, if evaluated by a Business & Management panel, may need to publish eight articles to ensure that at least four fall within the remit of ABS journals. The alternative, and arguably more likely scenario, is that IS researchers will change their publication patterns, shifting away from IDR and towards a more disciplinary focus.

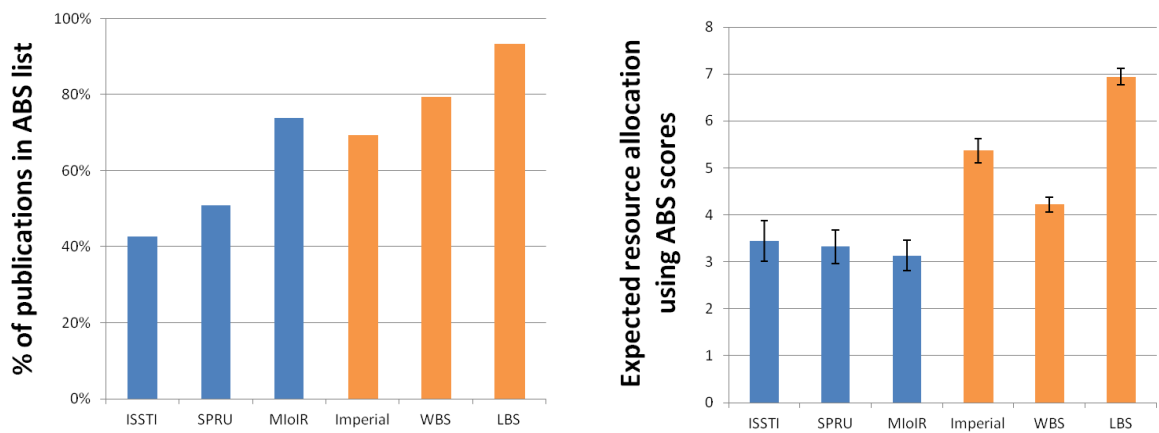

**Figure 10. Mechanisms of bias amplification: journal coverage and quasi-exponential scale in resource allocation.** Left: percentage of publications by unit indexed in the Web of Science (WoS) published in journals contained in the ABS list. Right: Expected outcome of resource allocation derived from an assessment exercise based on ABS journal scores (Figure 9) and a quasi-exponential scale.

A second mechanism amplifying the bias against IS is the exponential scale that the assessment exercise uses to reward perceived quality. In terms of (financial) resource allocation, this means that rank 1 articles have a multiplier of 0 (i.e. they are ignored), rank 2 articles have a multiplier of 1, rank 3 articles a multiplier of 3, and rank 4 articles a multiplier of 9. Using such a quasi-exponential scale, the 50% difference between MIoIR (Manchester) and LBS (London) (i.e. 1:1.5) in terms of performance as reflected in ABS journal scores would translate into a difference of 120% (i.e. 1:2.2) in the resources received by the units (see Figure 10). Given that this process has been cumulative over successive assessment exercises, there would be a strong incentive to shift publication practices, and therefore research, hiring and promotion patterns.

*5.2 Explanations for differences between Innovation Studies (IS) units and Business & Management Schools (BMS)*

---

[36] Business & Management is one of the Units of Analysis where there is also a high correlation (r=0.719) between the number of citations per paper and the 2001 RAE score (Mahdi et al., 2008, Table 3).

In principle, both IS and BMS might be expected to be equally interdisciplinary, as both deal with complex social issues. Business and management are not traditional scientific subjects and are to some degree multidisciplinary, with their research outputs published in economics, finance, and to a lesser degree psychology and operations research as well as business and management (though with little cross-linking). However, they are still much less interdisciplinary than IS units. Why are the knowledge base and audiences of IS units apparently so much more diverse than those for BMS?

Clausen et al.'s (this issue) survey of the drivers and barriers experienced by IS, Science and Technology Studies (STS) and entrepreneurship suggests IS and STS were established because of a 'need for cross-disciplinarity' and 'new academic knowledge'. By contrast, entrepreneurship units were created to establish a 'new academic teaching program' (ibid). If IS and STS centres were originally developed to carry out research in response to external policy and social questions, then it is little surprise that they are often driven simultaneously to engage with a wide range of stakeholders and diverse disciplines. By contrast, BMS may be more like entrepreneurship units, developing as centres of professional training and therefore primarily requiring a stock of scholars teaching in required fields, without any particular need for research integration. Senior BMS scholars have previously raised this point and expressed concern about the resulting inability of the field to address important social and managerial issues to a depth appropriate to their intrinsic complexity (Minzberg, 2000).

The problem-driven nature of nominal IS units may also explain why they are seldom 'purely' IS, as defined by studies of the core literature of IS. This putative IS core presumably lies in the middle left region of Figure 5, below the management and above the economics poles (see Table 6 in Fagerberg et al., this issue; Fagerberg and Verspagen, 2009). Instead, IS units tend to publish over a variety of IS-related fields, including the two management and economics poles, with an important presence in STS as shown in Figure 5, right middle area (see Table 4 in Martin et al.'s study of the core STS literature, this issue), as well as specific problem areas such as health, energy or environment.[37]

*5.3 How far can the findings be generalised?*

The field of Business & Management is perceived by some analysts as a rather paradoxical case in relation to other disciplines. Given that it is an applied field, one might expect to see a highly diverse knowledge base and a plurality of approaches. Instead, one finds BMS scholars competing to get published in a small number of very similar journals. This raises a question about the extent to which these findings on bias against IDR in BMS and IS are generalisable or only apply to these fields.

Research on journal rankings in economics (Oswald, 2007) suggests that the findings may at least be applicable to related social sciences. In many natural sciences the norm is to use indicators of journal quality, such as Thomson-Reuter's Impact Factor, rather than rankings. Could the use of Impact Factors discriminate against IDR? If IDR articles are less likely to be accepted in high Impact Factor journals, then this is clearly a possibility. According to the US National Academies (2004, p. 139):

> 'With the exception of a few leading general journals — such as *Science*, *Nature*, and the *Proceedings of the National Academy of Sciences* — the prestigious outlets for research scholars tend to be the high-impact, single discipline journals published by professional societies. Although the number of interdisciplinary journals is increasing, few have prestige and impact equivalent to those of single-discipline journals (…). Interdisciplinary researchers may find some recognition by publishing in single-discipline journals (…), but the truly integrated portion of their research may not be clear to much of the audience or be noticed by peers who do not read those journals.'

---

[37] This is perhaps an important difference between the core literature studies of IS and STS based on handbook chapters (Fagerberg et al., this issue; Martin et al., this issue), and the results obtained here based on our analysis of the journal publications from different research units. The former approach emphasises the theoretical foundations of IS and STS and their division, as shown, for example, in the relatively small degree of cross-citations (see Figure 2, Bhupatiraju et al., this issue; Leydesdorff, 2007b). A focus on organisational units, on the one hand, reveals their disparate intellectual debts and allegiances within the social sciences, and, on the other hand, their close engagement with practitioners in such areas as energy, biomedical research, health services and environment.

The correlation observed in Table 4 between the results based on ABS journal rankings and those based on the mean journal Impact Factor (after field normalisation) in comparison to the much lower correlation with the results based on using citations/paper may be interpreted as supporting the general hypothesis of an Impact Factor-based bias against IDR.

Similarly, although not all the Units of Assessment of the UK's assessment exercise were perceived as disadvantaging IDR departments, the possibility of such a bias has been repeatedly raised (e.g. Boddington and Coe, 1999), and it remains an open issue. As Martin and Whitley (2010, p. 64) noted:

> '…the UK has an essentially discipline- based assessment system for a world in which government policies are trying to encourage more user-focused and often interdisciplinary research. Those who have gone down the user-influenced route frequently conclude that they have ended up being penalized in the RAE process. (…) in practice the heavy reliance on peer review and the composition of RAE panels mean that discipline-focused research invariably tends to be regarded as higher quality.'

In summary, although in other fields the bias against IDR resulting from explicit or implicit perceptions of journal quality may not be as manifest or pronounced as in Business & Management, there are *a priori* grounds for believing that such a bias may nevertheless exist in any evaluation of IDR. However, further research is needed is order to test this suggestion, given the marked differences in the social institutionalisation between (and sometimes within) the various fields of natural science, engineering and social science (Whitley, 2000).

*5.4 The consequences of a bias against IDR*

We have so far argued that analyses based on ABS journal ratings may disadvantage IDR, and their use for evaluation purposes could therefore result in a bias against IDR that, in turn, may have significant financial repercussions under current REF procedures.[38] But what would the consequences be for society of such discrimination against IDR? A major intent behind both assessment and rankings is to foster competition (which is assumed to have desirable consequences) by providing fair, transparent, accountable procedures and rules by which this competition can be managed (Gläser and Laudel, 2007; pp. 108-109).[39] However, several analysts have warned against the 'inadvertent' but 'potentially destructive' consequences of bibliometric rankings for the science system (Weingart, 2005, p. 130; Roessner, 2000) and the capture of evaluation by disciplinary elites (Martin and Whitley, 2010, pp. 64-67).

---

[38] Note that the argument thus far has been based on a relatively naive understanding of indicators as simple measurement tools that can have unintended consequences if they are biased. A more politically-nuanced view on the role of indicators would also consider the performative power of journal rankings – namely, that rather than simply setting standards to help measure quality, they reflect deliberate aims to establish what that quality should be. The ABS journal rankings guide, for example, state that the function of the rankings is to '[p]rovide an indication of where best to publish' (ABS, 2010, p. 2).

A first type of consequence may be the creation or reinforcement of disincentives for researchers to engage in IDR. Among US sociologists and linguists, for example, Leahey (2007) found that more interdisciplinary (or less specialised) researchers tend to earn less.[40] Van Rijnsoever and Hessels (2011) reported that those researchers engaged in disciplinary collaboration benefit more in terms of promotion than those engaged in IDR. As noted previously, Lee and Harley have repeatedly argued that bias in the UK research assessment exercise has shifted recruitment in UK economics departments towards mainstream economists and away from heterodox economists (Harley and Lee, 1997; Lee and Harley, 1998; Lee, 2007). This push towards the disciplinary mainstream is also suggested by an analysis of the economics-related submissions in the new Italian research assessment exercise. Corsi et al. (2011) showed that the percentage of papers in heterodox economics and economic history was much lower in the assessment exercise than in a general economics database such as EconLit, suggesting a selection bias towards the more narrowly disciplinary specialties within economics, such as econometrics and finance.

Second, the bias may stimulate a process of intellectual inbreeding, where efforts to increase the quality of research end up creating a self-reinforcing narrowing down of conceptions of quality, ultimately affecting the very content of research (Mirowski, 2011). Ultimate responsibility for the definition of quality may shift from the disciplinary elite to the audit process itself. A number of prominent management scholars have expressed concerns that this is already happening, and that some parts of management research are becoming an irrelevant game structured by the academic job market and business school rankings rather than by research excellence or concern about real-world issues (Minzberg, 2000; Willmott, 2011a, 2011b; Alvesson and Sandberg, 2011; Tourish, in press).[41]

Thirdly, since socially relevant research almost inevitably involves navigating across or between several disciplines, a reduction in IDR may shift the orientation of research away from complex social questions. For example, a study by Goodall (2008) reported that 'over the past few decades, only 9 articles on global warming or climate change have appeared in the top 30 Business & Management titles', out of approximately 31,000 papers in total (p. 417). By contrast, more than 2,000 publications had appeared on the topic in 'journals that are peripheral to the main social science disciplines' (p. 415). Goodall (2008) attributes this dearth of publications on climate change in top journals to their valuing of theory over practical issues, political bias and associated career incentives. Patenaude (2011) recently confirmed Goodall's findings and showed that this lack of research interest may spill over into teaching, as MBA curricula also display relatively limited interest in climate change (p.

---

[39] This is based on the assumption that academics are indeed seeking to 'win' some form of competition with each other, either as individuals or as departments, rather than being engaged in a shared, international, cumulative, intellectual endeavour.

[40] Interestingly, this may partly explain earning differences between men and women, given that women tend to be more interdisciplinary (Leahey, 2007; Rhoten and Pfirman, 2007; Van Rijnsoever and Hessels, 2011).

[41] Further evidence of these concerns comes from the special symposium at the EGOS 2011conference organised by Alvesson and Sandberg to debate the issue with the editors of the *Journal of Management Studies* and *Organization Studies*, and from the recent publication of a special issue of *Organization* (see Willmott, 2011a and the ensuing papers).

260). Patenaude partly attributes such a bias to corporate values and beliefs as well as to existing academic incentives and communication channels.

Lastly, this bias against IDR may reduce the cognitive diversity of the entire science system. Diversity in the science system is important from an ecological or evolutionary perspective because: (i) notions of quality are dynamic (what is marginal now may later become highly significant); (ii) diversity helps prevent paradigmatic lock-in; and (iii) diversity fosters the appearance of new types of knowledge (Stirling, 1998, pp. 6-36; Stirling, 2007). All in all, there are good reasons to be concerned about the findings in this paper.

The problems associated with the current focus on performance evaluation, in the UK and elsewhere, cannot be dealt by minor changes. Instead, they will require a fundamental re-think of the goals of research assessment given the systemic nature of scientific development. Minor reforms could improve existing evaluation exercises by recognising the uncertainties involved in evaluation and tackling some of the issues raised. For example: requiring the declaration of standard errors or statistical significance; using contrasting and more sophisticated normalisations (such as the fractional citation measure proposed); and adopting mathematically rigorous representations of citation distributions instead of means (Leydesdorff and Bornmann, 2011). However, these changes do not address the key flaw in the current system which is the assumption that maximisation of the individual unit's 'performance' improves overall systemic performance. An alternative view is that science is an open, extended and complex system with a range of competing (and legitimate) perceptions of performance. This implies a radically different approach to evaluation and funding focusing on as much on the prospective fostering of potentially fruitful future integrations and the realising of possible strategic synergies, as on retrospective attributions of narrow notions of past 'success' (Molas-Gallart and Salter, 2002; Klavans, personal communication, 7$^{th}$ November 2011).

**6. Conclusions**

This empirical investigation has responded to wider concerns that have been raised in science policy debates about the evaluation of IDR. It has involved a more rigorously 'plural and conditional' approach to research evaluation, making use of a number of 'converging partial indicators'. Using a range of innovative maps and metrics, the paper has confirmed that IS units are indeed more interdisciplinary than leading BMS when viewed under various perspectives. More importantly, it has shown that the widespread use of ABS journal rankings in BMS results in a bias in favour of disciplinary research, while conversely the research of interdisciplinary IS units tends to be assessed as being of lower quality. However, that lower assessment is not supported by citation-based indicators, which are generally considered to offer more robust measures of performance. Consequently, the study suggests that the use of ABS journal rankings systematically disadvantages IDR in this setting. This finding clearly needs to be tested in a wider context, in particular in the natural sciences, in order to establish its robustness and the extent to which the problems identified here are generalisable. The main caveats are that citation data were collected for only a relatively short period after publication, without using a fixed 'citation window'; and that we used conventional, mean-based performance measures instead of more advanced, distribution-based measures (Leydesdorff and Bormann, 2011).

These quantitative findings support what is by now a fairly well-established picture, evident from qualitative investigations in science studies and in science policy (National Academies, 2004), that criteria of excellence in academia are essentially based on disciplinary standards, and that this hinders interdisciplinary endeavours in general, and policy and socially relevant research in particular (Travis and Collins, 1991; Langfeldt, 2006). Few previous studies have investigated bias against IDR in quantitative terms (Porter and Rossini, 1985; Rinia et al., 2001a). Consequently, this study apparently constitutes one of the first and most thorough explorations of whether one of the criteria most widely used in research assessment, namely journal rankings, may result in a bias against interdisciplinary research. We find strong evidence that it does.

In recent decades criteria of quality have become institutionalised in the form of rankings that can have major (and often negative) reputational and funding implications. The use of a simple ranking procedure is predicated on the assumption that the results constitute *objective assessments* that can be treated as robust proxies for academic excellence. The empirical results in this paper challenge such claims to objectivity. They suggest instead that such an approach generates a rather narrow and idiosyncratic view of excellence. To the extent that ABS-style journal rankings are increasingly used to evaluate individual and organisational research performance, it does seem possible to identify a *prima facie* hypothesis that this practice exercises a suppressive effect on IDR.[42]

In summary, this paper has shown that when journal rankings are used to help determine the allocation of esteem and resources, they can suppress forms of interdisciplinarity that are otherwise widely acknowledged to be academically and socially useful. Important implications arise, both for research evaluation in the specific fields in question, as well as for wider investigations to inform the more general governance of science and technology using metrics to capture multidimensional qualities that cannot be intrinsically reduced to a single indicator.

**Supplementary materials**

The full suite of maps (diversity, coherence and intermediation) for each unit and perspective (publications, references and citations) is available at http://www.interdisciplinaryscience.net/maps and in the supplementary files, in PowerPoint format:

- SupplementaryFile1: Disciplinary Diversity (subsection 4.1)
  http://www.sussex.ac.uk/Users/ir28/IDR/Disciplinary_Diversity.pptx
- SupplementaryFile2: Disciplinary Coherence (subsection 4.1)
  http://www.sussex.ac.uk/Users/ir28/IDR/Disciplinary_Coherence.pptx
- SupplementaryFile3: Intermediation (subsection 4.1)
  http://www.sussex.ac.uk/Users/ir28/IDR/Intermediation.pptx
- SupplementaryFile4: Diversity ABS Rankings (subsection 4.2)
  http://www.sussex.ac.uk/Users/ir28/IDR/ABS_Ranking_Diversity.pptx
- SupplementaryFile5: Comparison of Units' Performances (subsection 4.3)
  http://www.sussex.ac.uk/Users/ir28/IDR/Performance_Comparison.pptx

---

[42] This study used conservative, internal measures of scientific performance. It is likely that a more thorough evaluation that took into account efficiency (i.e. performance given costs) or broader social impact would strengthen the findings on bias.

## Acknowledgments

We thank colleagues from various British Business & Management schools for feedback. We are grateful to Jan Fagerberg, Ben Martin and three anonymous referees for detailed comments and Diego Chavarro for designing the website www.interdisciplinaryscience.net. IR and AO were funded by the US NSF (Award no. 0830207, http://idr.gatech.edu/) and the EU FP7 project FRIDA (grant 225546, http://www.fridaproject.eu). The findings and observations contained in this paper are those of the authors and do not necessarily reflect the views of the funders. Loet Leydesdorff acknowledges Thomson Reuters for the use of Journal Citation Reports data.